%% file: main.tex
\documentclass[lettersize,journal]{IEEEtran}
\usepackage{amsmath,amsfonts}
\usepackage{algorithmic}
\usepackage{algorithm}
\usepackage{array}
\usepackage[caption=false,font=normalsize,labelfont=sf,textfont=sf]{subfig}
\usepackage{enumitem}
\usepackage{textcomp}
\usepackage{stfloats}
\usepackage{hyperref}
\usepackage{cleveref}
\usepackage{url}
\usepackage{verbatim}
\usepackage{graphicx} 
\usepackage{subcaption}  
\usepackage{cite}
\usepackage{xcolor}
\graphicspath{{figs/}{figures/}{pictures/}{images/}{./}} 
\newcommand{\revise}[1]{{\textcolor{black}{#1}}}
\newcommand{\change}[1]{{\textcolor{black}{#1}}}

\begin{document}

\title{Hairpin Vortices Extraction in Turbulent Boundary Layer Flows}

\author{Adeel~Zafar, Zahra~Poorshayegh, Lei~Si, Di~Yang, Guoning~Chen%
\thanks{\textbullet\ Adeel~Zafar, Lei~Si, and Guoning~Chen are with the Department of Computer Science, University of Houston, Houston, TX, USA.}%
\thanks{\textbullet\ Zahra~Poorshayegh and Di~Yang are with the Department of Mechanical and Aerospace Engineering, University of Houston, Houston, TX, USA.}%
\thanks{\textbullet\ E-mail: azafar3@uh.edu, zpoorsha@cougarnet.uh.edu, lsi@uh.edu, diyang@uh.edu, gchen22@central.uh.edu.}%
}

\markboth{Accepted in IEEE Transactions on Visualization and Computer Graphics}%
{}


\maketitle
\begin{abstract}
\input{content/abstract}
\end{abstract}


\input{content/introduction}

\input{content/related_works}
\input{content/method}
\input{content/results}

\input{content/discussion}

\section*{Acknowledgments}
This research was supported by NSF OAC 2102761.

\bibliographystyle{IEEEtran}
\bibliography{main}

\begin{IEEEbiography}[{\includegraphics[width=1in,height=1.25in,clip,keepaspectratio]{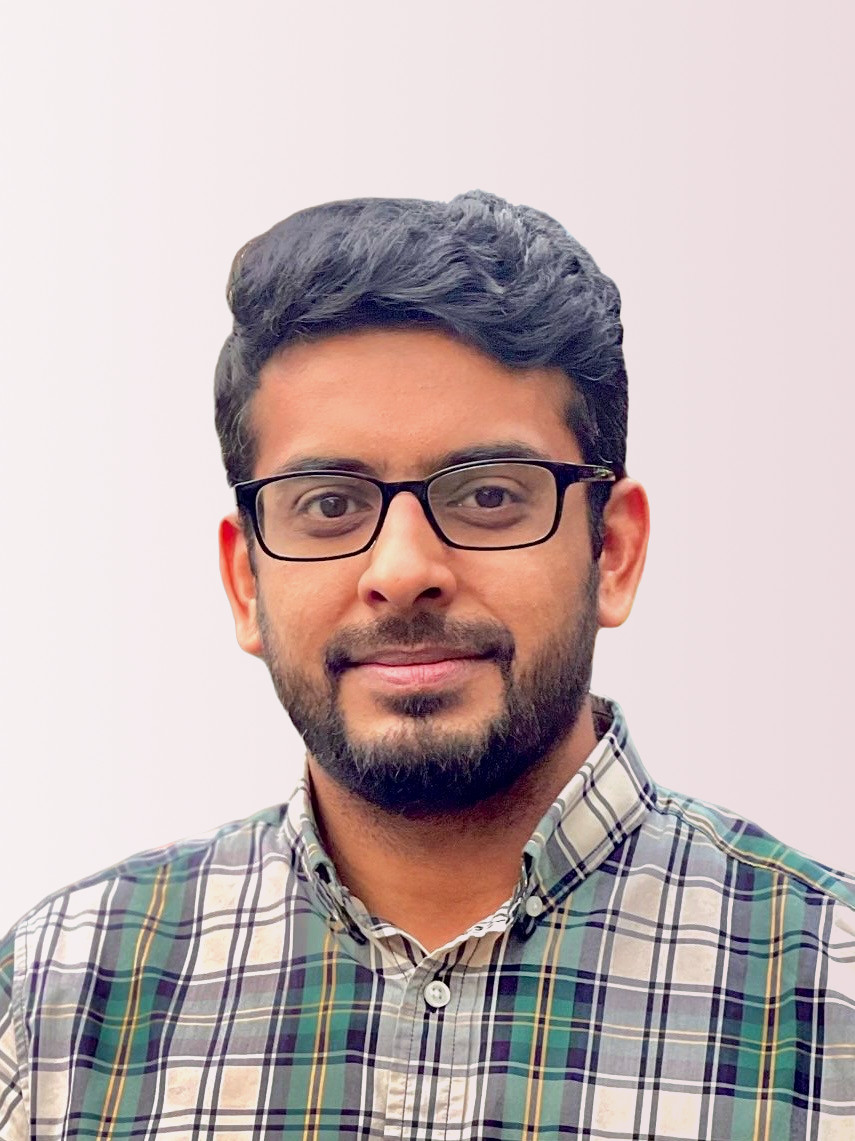}}]
{Adeel Zafar}
is a Ph.D. student in the Department of Computer Science at the University of Houston. He earned an M.S. in Software Engineering from Shanghai Jiao Tong University in 2019 and a B.E. in Computer Software Engineering from the National University of Sciences and Technology, Islamabad, in 2015. His research interests include visualization, data analytics, deep learning, and high-performance computing.
\end{IEEEbiography}

\begin{IEEEbiography}[{\includegraphics[width=1in,height=1.25in,clip,keepaspectratio]{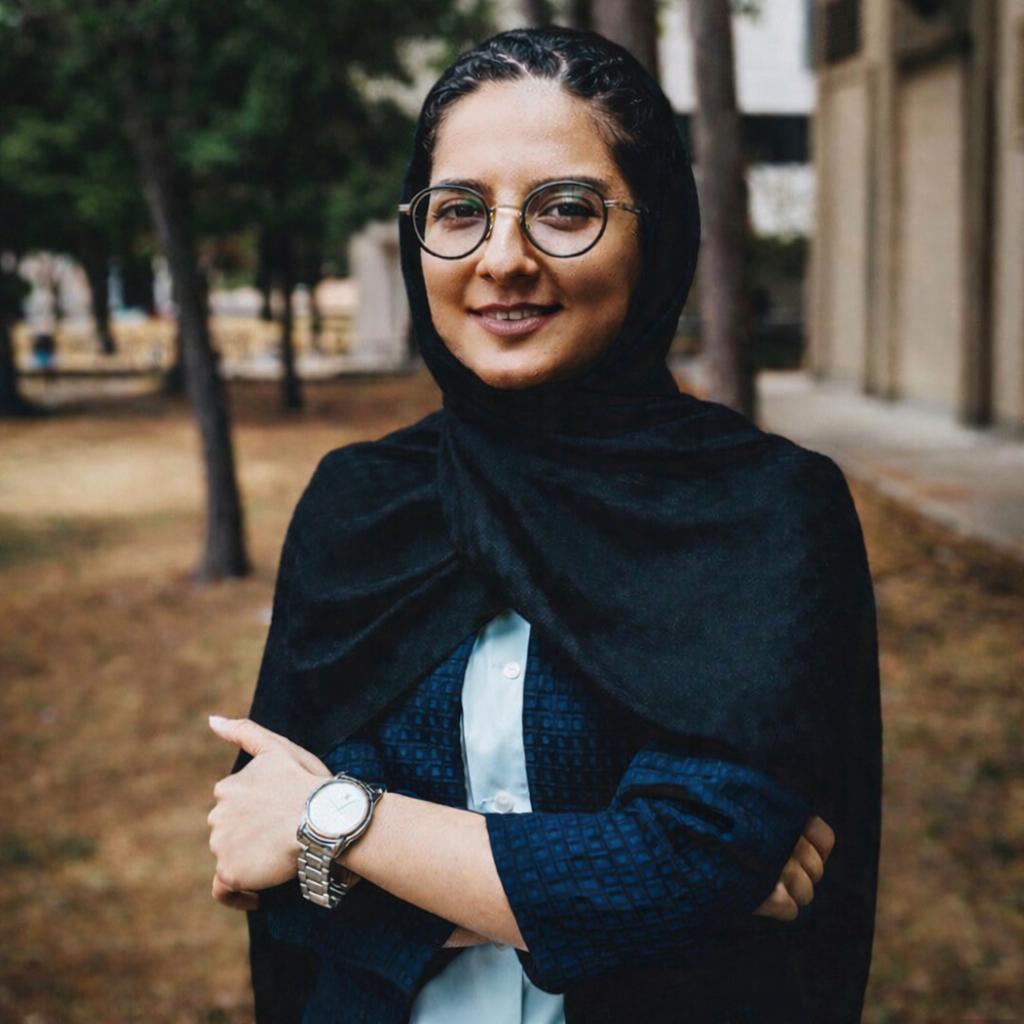}}]
{Zahra Poorshayegh}
 is a Ph.D. student in the Department of Mechanical and Aerospace Engineering at the University of Houston. She received her B.S. and M.S. degrees from Sharif University of Technology. Her research interests include computational fluid dynamics, turbulence, and vortex dynamics. Specifically, she works on wall-bounded turbulence, using computational fluid dynamics and data-driven analysis to study coherent structures in turbulent flows.
\end{IEEEbiography}

\begin{IEEEbiography}
[{\includegraphics[width=1in,height=1.25in,clip,keepaspectratio]{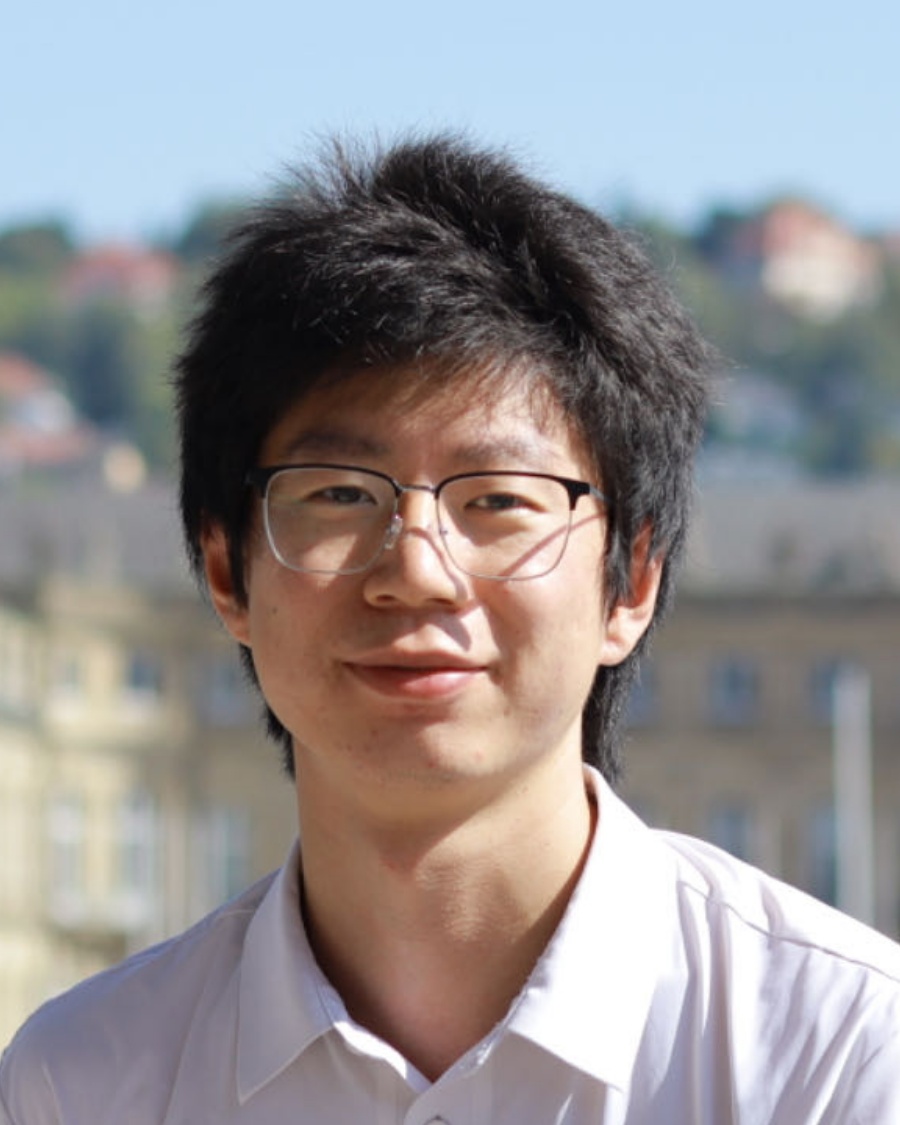}}]
{Lei Si}
 received a Ph.D degree in Computer Science from the University of Houston in 2025, an M.S. degree with honors in Computer Science from the University of Illinois at Springfield in 2020, and B.E. in Cybersecurity at North China University of Technology in 2018. His research focuses on geometric modeling, visualization, physically-based simulation, computer vision, cyber security, and artificial intelligence.
\end{IEEEbiography}

\begin{IEEEbiography}[{\includegraphics[width=1in,height=1.25in,clip,keepaspectratio]{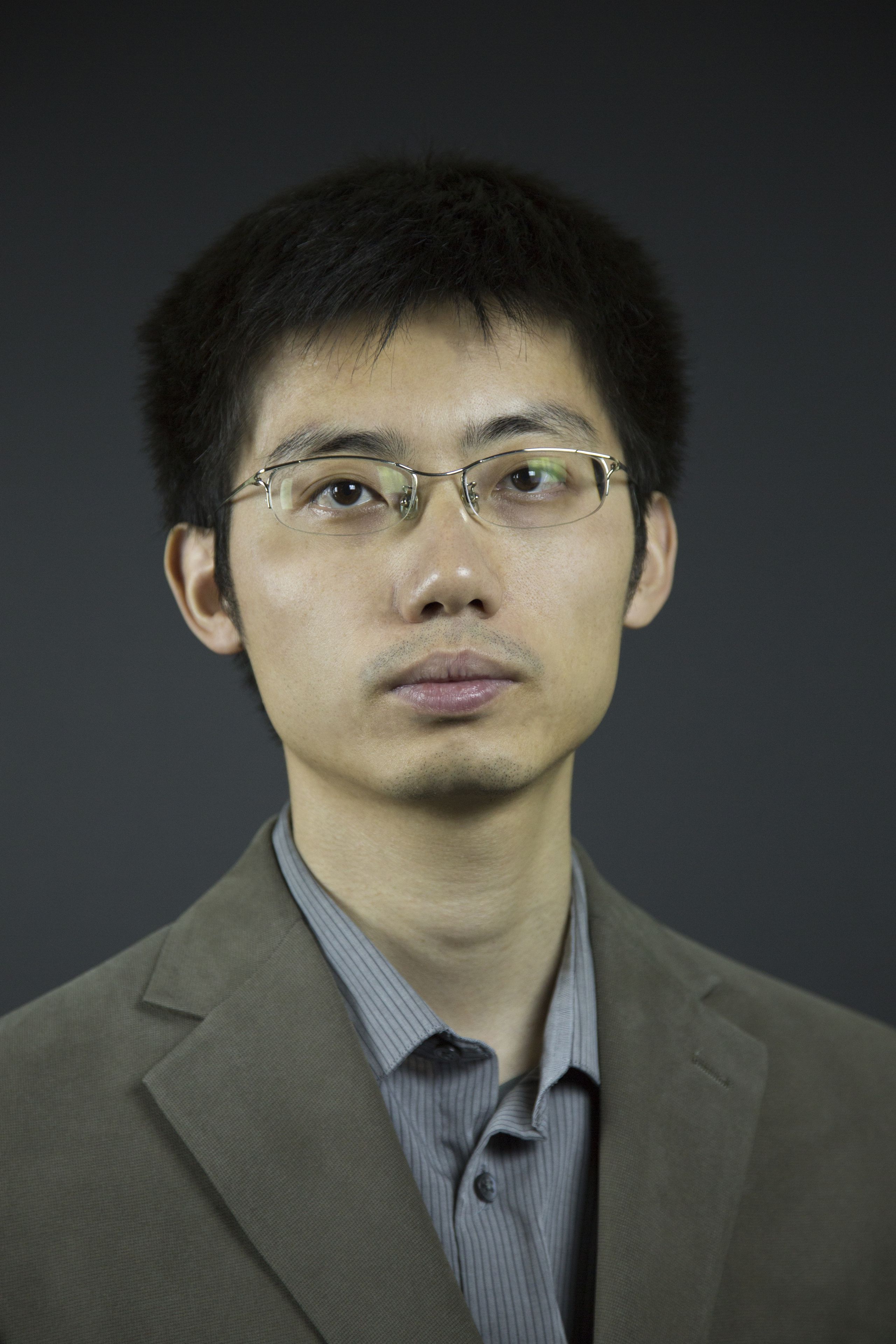}}]
{Di Yang}
is an Associate Professor in the Department of Mechanical and Aerospace Engineering at University of Houston. He earned his B.S. from the University of Science and Technology of China and completed his M.S. and Ph.D. at Johns Hopkins University. His research focuses on hydrodynamic turbulence, using high-performance computing to study complex flows, with interests in simulations of turbulence, wind energy systems, ocean dynamics, and oil spill dispersion.
\end{IEEEbiography}

\begin{IEEEbiography}[{\includegraphics[width=1in,height=1.25in,clip,keepaspectratio]{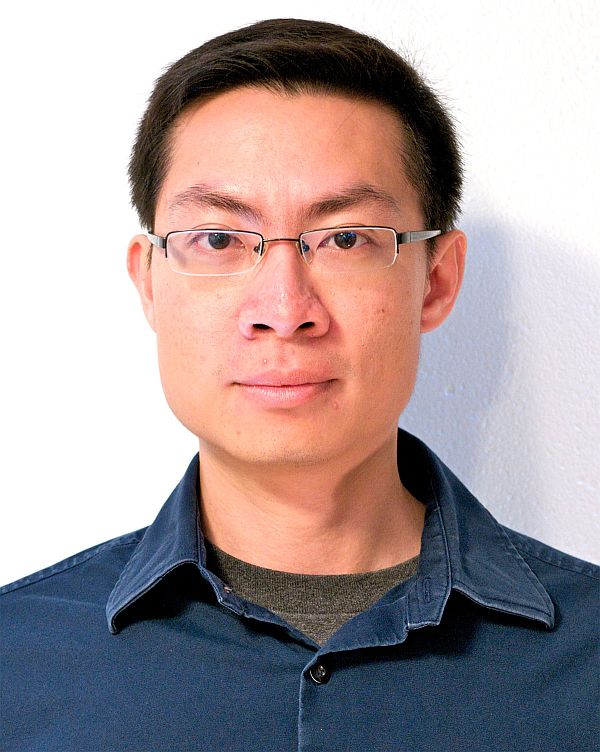}}]{Guoning Chen}
is a Professor at the Department of Computer Science at the University of Houston. He earned a Ph.D. degree in Computer Science from Oregon State University in 2009. His research interests include visualization, data analytics, computational topology, geometric modeling, geometry processing, and physically-based simulation. He is a senior member of IEEE and a member of ACM.
\end{IEEEbiography}
\end{document}

%% file: content/abstract.tex
Hairpin vortices are fundamental structures within turbulent boundary layers, playing a crucial role in energy dissipation, mixing, and momentum transport. However, accurately extracting these structures remains challenging due to their irregular shapes, varying scales, and entanglement with surrounding vortical structures. This paper presents a novel framework for the extraction of hairpin vortices from turbulent boundary layers. The method begins by identifying vortical regions and decomposing them into smaller segments using merge tree–based segmentation. A novel bottom-up rejoining approach is then introduced to group candidate segments according to the geometric and physical characteristics of hairpin vortices, resulting in regions that encompass complete hairpin vortex structures. These regions are subsequently refined and validated through skeleton analysis to detect the characteristic hairpin shape and are further confirmed using additional scalar-based criteria. Finally, smooth enclosing surfaces are generated for effective visualization. To enable quantitative evaluation, reference hairpin vortices are extracted from several flow datasets and used as ground truth. Compared with existing approaches, the proposed method eliminates manual parameter tuning, reduces under- and over-segmentation, and significantly improves both accuracy and computational efficiency. Demonstrations on multiple turbulent flow cases show that the method is robust and effective for hairpin vortex extraction under varying boundary layer conditions.

%% file: content/introduction.tex
\section{Introduction}
\label{sec:intro}
A boundary layer is the thin region of fluid immediately adjacent to a solid surface where fluid motion is influenced by surface friction, resulting in velocity shear that generates and sustains turbulence. Hairpin vortices are widely recognized as fundamental coherent structures within turbulent boundary layers, where they play a critical role in influencing the dynamics of turbulent flow. These vortices are instrumental in enhancing momentum transfer between different layers of the flow, thereby promoting efficient mixing processes and significantly contributing to the dissipation of turbulent kinetic energy~\cite{ADRIAN_MEINHART_2000, adrian2007hairpin}. Characterized by their distinctive `hairpin' shape, which features a curved head region, elevated necks, and trailing legs extending upstream, these structures have been extensively observed in both experimental studies \cite{hutchins2014hairpin} and high-fidelity numerical simulations \cite{zhou1999mechanisms} of wall-bounded turbulence. The presence of hairpin vortices is closely linked to the generation of near-wall low-speed streaks and bursting events, which are key mechanisms sustaining turbulence in boundary layers~\cite{zhou1999mechanisms}. 
\change{Accurate identification and extraction of hairpin vortices are essential for advancing the understanding of turbulent flow physics. From a visualization perspective, precise extraction is a prerequisite for generating clear, unobstructed representations of coherent structures. By isolating these structures, the proposed method mitigates visual clutter and spatial occlusions, enabling fluid dynamics experts to analyze shapes, spatial organization, and temporal evolution with high structural clarity. However, the irregular and entangled nature of hairpin vortices remains a significant challenge for intuitive visual exploration.}

\begin{figure}[t]
 \centering
    \includegraphics[width=0.99\linewidth]{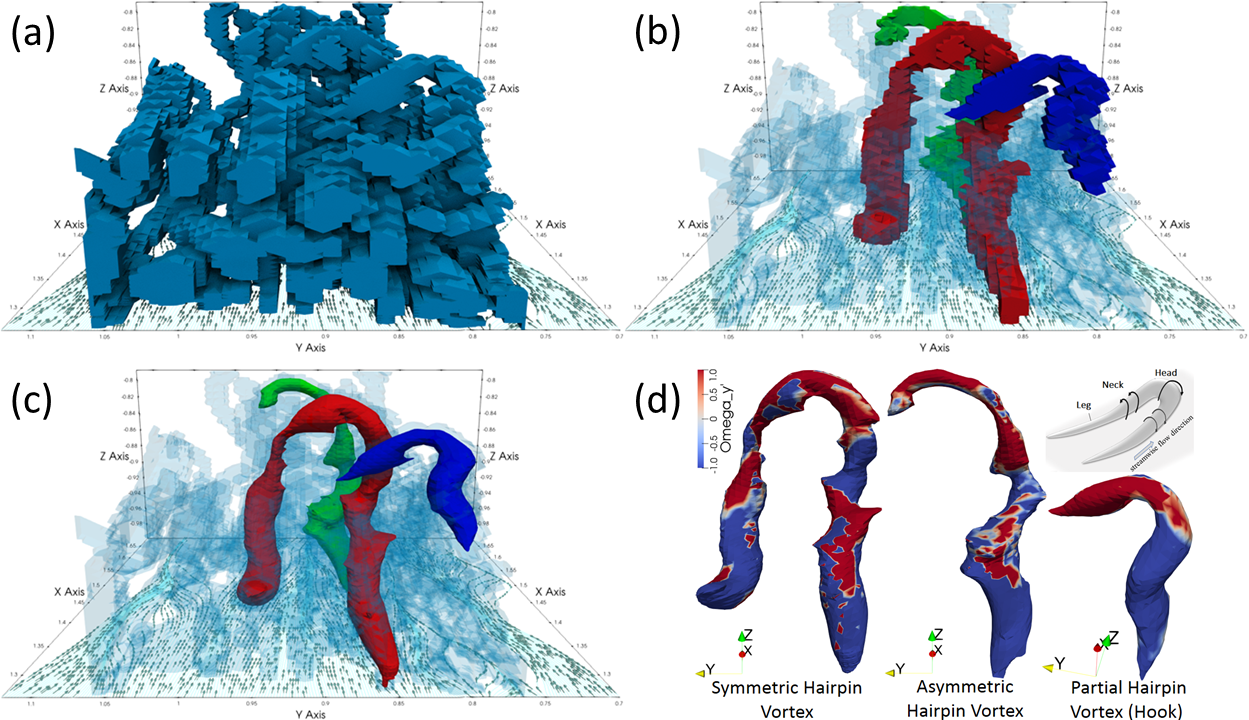}
    \caption{(a) Multiple overlapping vortices in a single vortical region. (b) The fully separated hairpin vortices. (c) A smooth visualization of the identified hairpin vortices. (d) Different possible types of the identified hairpin vortices, color-coded by spanwise vorticity fluctuation $\omega_y'$, in comparison with the ideal shape. It can clearly be seen that the hairpin vortices are embedded within the vortical regions while overlapping with other vortices and exhibiting complex shape configurations, making their identification a challenging problem.}
 \label{fig:1_1}
\end{figure}

One common approach for vortex identification in fluid flows relies on scalar criteria, such as $\lambda_2$~\cite{jeong1995identification}, $Q$~\cite{hunt1987vorticity}, and $\lambda_{ci}$~\cite{zhou1999mechanisms}, which are widely used to detect regions of strong vortical flow motions. These scalar-based methods are effective for initial vortex detection, as they highlight areas of high vorticity or swirling motions. However, they are insufficient for accurately isolating specific structures such as hairpin vortices. Hairpin vortices are typically embedded in highly complex, interconnected vortical regions where multiple structures overlap and interact, as shown in \cref{fig:1_1}. This complexity makes it difficult to distinguish the characteristic hairpin shape using scalar criteria alone. Moreover, hairpin vortices are defined not only by swirling motion, but also by their distinct geometry, including a curved head, vertical necks, and trailing legs, which conventional criteria do not automatically capture. As a result, more detailed segmentation and refinement beyond simple scalar thresholds are required to accurately identify, extract, and verify hairpin vortices in turbulent boundary layers, enabling a clearer understanding of their role in flow dynamics and turbulence generation.

Zafar et al.~\cite{adeelhairpin2023} introduced a framework for extracting hairpin vortices by first identifying general vortical regions using a $\lambda_2$-based region-growing approach, followed by segmentation via progressive extraction of $\lambda_2$-based isosurfaces. After segmentation, their approach combines geometric characteristics (e.g., curvature) with physical properties (e.g., spanwise vorticity fluctuation $\omega_{y}'$) of segmented vortices to identify candidate hairpin structures. A subsequent clustering step is then employed to separate true hairpin vortices from false positives. \revise{However, as detailed in \cref{sec:background}, this method suffers from both under- and over-segmentation. Over-segmentation causes hairpin vortices to split unnecessarily, failing to retain the whole structure intact, whereas under-segmentation causes nearby vortices to remain attached to the hairpin vortices, thus failing to resolve the vortex entanglement problem.} Moreover, their method requires considerable parameter tuning and a user feedback loop. Zafar et al.~\cite{zafar2024topological} partially addressed the over-segmentation issue by combining merge tree-based segmentation of vortical regions with vortex lines. However, as discussed in \cref{sec:background}, this approach does not generalize well to handle complex vortical regions and remains computationally expensive (\cref{tab:Table1}).

In this study, we propose to address the issues of under- and over-segmentation in hairpin vortex extraction by a more effective and efficient strategy, which integrates topology-based segmentation of vortical regions with a bottom-up rejoining step guided by physical and geometric criteria. Our new approach begins by extracting initial vortical regions using the $\lambda_2$-criterion (as in~\cite{adeelhairpin2023}), followed by a merge tree–based segmentation~\cite{carr2003computing} to decompose these regions into smaller segments (similar to, but different from,~\cite{zafar2024topological}). We then introduce a rejoining stage that first identifies segments corresponding to hairpin vortex heads using the $\omega_{y}'$ criterion. Using these head segments as seeds, we compute vortex lines and merge overlapping segments to form candidate regions that encapsulate entire hairpin structures. The candidate regions are subsequently refined and validated by analyzing their skeletons, detecting the characteristic hairpin geometry, and enforcing physical validation through additional scalar-based criteria. Finally, the validated hairpin vortices are visualized using smooth enclosing surfaces. To enable quantitative and qualitative evaluations, we provide the first set of reference hairpin vortices for multiple turbulent flow datasets (Couette flow~\cite{li2019direct}, channel flow~\cite{lee2013petascale}, and transitional boundary layer~\cite{Li2008TurbulenceDB}). Our approach demonstrates higher accuracy, better quality, improved computational efficiency, full automation and enhanced generalization compared to previous techniques, making it a practical solution for extracting hairpin vortices in turbulent boundary layers. \change{The code can be found at \textit{https://github.com/adeelz92/hairpin-extraction-tvcg.git}.}


%% file: content/related_works.tex
\section{Background}
\label{sec:background}
In this section, we introduce the key technical terms used throughout the paper, review related work on general vortex extraction, and discuss two closely related methods for hairpin vortex extraction.

\subsection{$\lambda_2$ Criteria and Vorticity}
\noindent \change{In this work, we use $\lambda_2$ as the primary scalar field for extracting vortical regions because it is Galilean invariant~\cite{gunther2018state}, ensuring that identified structures are independent of the observer’s translational velocity or the underlying mean flow. Previous comparisons have shown that $\lambda_2$-based extraction is preferred over vorticity ($\omega$) because the latter tends to yield distorted, 'pancake-like' structures in shear layers rather than the distinct, tube-shaped structures characteristic of coherent vortices~\cite{bremer2015identifying}. Vorticity ($\omega$) is used only for the refinement steps in our pipeline (\cref{sec:method})}. Given a flow velocity vector field $\vec{u}=(u,v,w)$, where $u$, $v$ and $w$ are the velocity components in the $x$, $y$ and $z$ directions, respectively, the velocity gradient tensor is defined and decomposed as $\nabla \vec{u} = \boldsymbol{S} + \boldsymbol{\Omega}$, where $\boldsymbol{S} = \frac{1}{2} \left( \nabla \vec{u} + (\nabla \vec{u})^\mathrm{T} \right)$ 
and $\boldsymbol{\Omega} = \frac{1}{2} \left( \nabla \vec{u} - (\nabla \vec{u})^\mathrm{T} \right)$ are the symmetric and antisymmetric parts of the velocity gradient tensor, respectively.  The $\lambda_2$ criterion defines vortical regions as areas where the second largest eigenvalue of the tensor $\boldsymbol{S}^2 + \boldsymbol{\Omega}^2$ is negative. That is, a region belongs to a vortex if $\lambda_2 < 0$. In practice, using $\lambda_2 < 0$ is often not restrictive enough, so an appropriate threshold must be chosen to accurately extract vortices. The vorticity vector is defined as $\vec{\omega} = \nabla \times \vec{u}$. The spanwise vorticity component is given by $\omega_y = \frac{\partial u}{\partial z} - \frac{\partial w}{\partial x}$. The spanwise vorticity fluctuation is obtained by decomposing $\omega_y$ into its mean and fluctuating parts: $\omega_y' = \omega_y - \overline{\omega_y}$, where $\overline{\omega_y}$ denotes the horizontally averaged spanwise vorticity.

\subsection{Merge trees}
\label{subsec:mergetrees}
\begin{figure}[!t]
 \centering
    \includegraphics[width=0.99\linewidth]{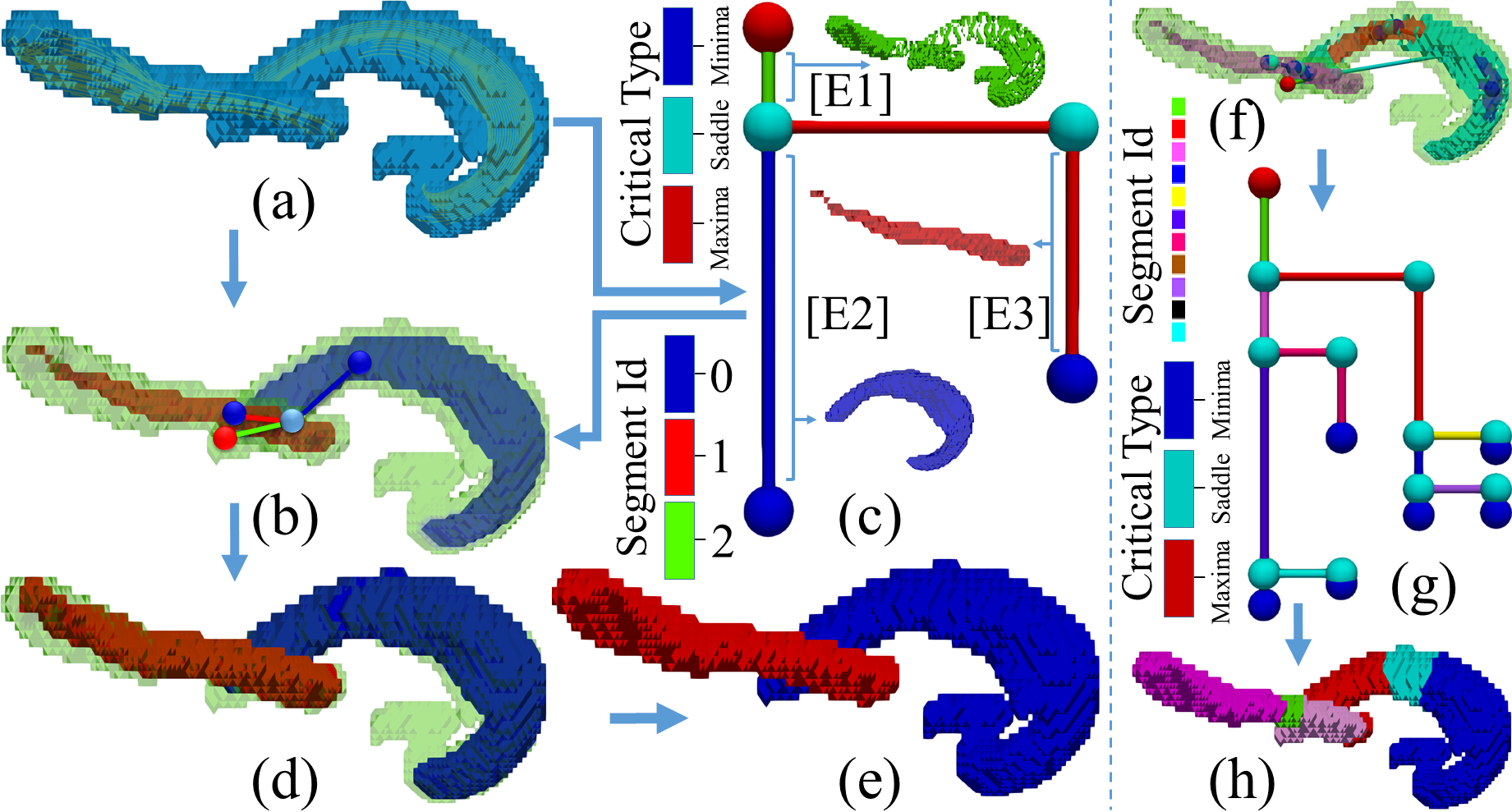}
    \caption{This figure illustrates the merge tree-based segmentation process. (a) A $\lambda_2$-based vortical region. (b) Its segmentation using a minimal merge tree consisting of a maximum (red sphere), a saddle (cyan sphere), and two minima (blue spheres). (c) The planar view of the minimal merge tree. The saddle-minima edge pairs (E2, E3) correspond to the leaf segments (blue and red in (b)), which are surrounded by the parent segment (green in (b)) corresponding to the saddle-maximum edge pair (E1). A merge tree may contain many critical points, leading to hierarchical segmentation, as shown in (g) and (f). (b), (d), and (e) illustrate the layering process, which converts the hierarchical segmentation in (b) into the linear segmentation in (e). (h) shows the layering results using the full merge tree-based segmentation of the region in (f).}
 \label{fig:3_1A}
\end{figure}

Merge trees~\cite{carr2003computing} are topological constructs that encode the merging relationships among the level sets of scalar fields. For a scalar field $f(p) \in \mathbf{R}$ defined over a set of points in 3D space, $p \in \mathbf{R}^3$, a level set is defined as $\{p \in \mathbf{R}^3 \ | \ f(p) = c\}$, representing the set of points where $f(p)$ equals a constant value $c$. As the value of $c$ increases, these level sets merge, and such transitions are captured by merge trees. The evolution of level sets occurs at \textit{critical points}, which are points where the gradient of the scalar field, $\nabla f(p) = \mathbf{0}$. The level sets begin (take birth) at minima, as the value of $f(p)$ increases, level sets merge at saddle points and vanish (die) at maxima. The extent to which a level set stays alive is encoded by persistence~\cite{edelsbrunner2022computational}. Consider a region that consists of only two minima, a saddle, and a maximum. Its merge tree will form a minimal merge tree as shown in \cref{fig:3_1A}(b) and its planar view in \cref{fig:3_1A}(c). Minima form the leaf nodes of the tree (blue spheres in \cref{fig:3_1A}(d)). As the value of $f(p)$ increases, level sets (blue and red segments in \cref{fig:3_1A}(b)) merge at saddle points (cyan sphere in \cref{fig:3_1A}(d)). A parent level set (green segment in \cref{fig:3_1A}(b)) surrounds the child level sets (blue and red segments in \cref{fig:3_1A}(c)). This results in a hierarchical segmentation of the region (\cref{fig:3_1A}(b)). In practice, a scalar field may contain many critical points, resulting in a merge tree with multiple hierarchical levels (\cref{fig:3_1A}(f, g)).

\subsection{Related Work}
\label{sec:related}
Vortex extraction is essential in analyzing flow data \cite{gunther2018state}. Many region-based methods have been proposed to identify vortical structures, such as \( \lambda_2 \)~\cite{jeong1995identification}, \( Q \)~\cite{hunt1987vorticity}, \( \lambda_{ci} \)~\cite{zhou1999mechanisms}, \( \Delta \)~\cite{chong1990general}, and \( Rortex \)~\cite{liu2018rortex} criteria, as well as other approaches based on pressure \cite{hunt1988eddies} and vorticity. These methods rely on scalar quantities to characterize rotational behavior in fluid flows. While effective in many cases, region-based methods often require selecting thresholds to reveal meaningful and interpretable structures, which can be non-trivial and ad-hoc. Alternatively, line-based methods \cite{sujudi1995identification,PEI99,weinkauf07c,SchafhitzelVGWCE08,guo2021exact,rautek2023vortex} have been introduced, focusing on the extraction of vortex corelines. Vortex corelines represent the centers of rotational motion, and the Parallel Vector operator \cite{PEI99} is often used to detect them. However, due to the need to calculate high-order derivatives, these methods are often numerically unstable, resulting in disconnected coreline segments that are difficult to correct~\cite{adeel2022hairpin}.  

Vortices can also be detected using geometric-based methods \cite{sadarjoen1999geometric,sadarjoen2000detection}, integration-based techniques \cite{wiebel2011topological,weinkauf2010streak}, and objective methods \cite{gunther2018state}. Moreover, vortex behaviors can be revealed by studying the temporal evolution of attributes along pathlines \cite{berenjkoub2018visual,nguyen2021physics}. Techniques have been proposed to detect vortex boundaries to support quantitative studies of vortices \cite{haller2016IVD,berenjkoub2020vortex}. Recently, machine learning (ML) approaches have been applied to general vortex detection \cite{deng2019cnn,wang2020rapid}, ocean eddy (2D vortex) detection \cite{lguensat2018eddynet,franz2018ocean,Bai19:OceanEddyCNN}, vortex boundary identification \cite{berenjkoub2020vortex}, and vortex coreline extraction \cite{Kim19EuroVis}.

Region-based methods remain widely used in fluid mechanics due to their simplicity and computational efficiency, with various specialized techniques developed for specific turbulent flows~\cite{nguyen2020Taylor}. However, a major challenge is the under- or over-segmentation of vortices, where extracted regions either mistakenly merge multiple structures or fail to capture a vortex in its entirety. This issue stems from the fact that optimal scalar thresholds are rarely known a priori and vary across datasets. Even within a single flow, vortical structures exhibit varying intensities of swirling motion due to local physics, making a universal threshold ineffective. Consequently, to identify specific types like hairpin vortices, they must first be isolated from these complex, overlapping regions.

Topological approaches, such as merge trees, have emerged as a powerful tool for flow analysis by capturing the hierarchical structure of scalar fields. This enables the structured decomposition of complex, entangled vortical regions and supports multiscale vortex separation and tracking. Laney et al.~\cite{laney2006understanding} utilized Morse theory to analyze Rayleigh–Taylor instabilities through hierarchical bubble segmentation, while Bremer et al.~\cite{bremer2015identifying} introduced topological segmentation of indicator functions to extract regions without relying on global thresholds. Further advancing this, Qu et al.~\cite{qu2024vortex} linked velocity-gradient eigenvalues to vortex stability and swirling strength, and Bridel-Bertomeu et al.~\cite{bridel2019topological} applied TDA to track eddy evolution in high-velocity compressible flows. Recently, Zhu et al.~\cite{Zhu_Xi_2019} proposed VATIP to classify vortices, such as quasi-streamwise, hairpins, hooks, and branches, by analyzing axis-line topology and clustering. However, identifying individual vortices remains challenging at high Reynolds numbers, where vortical structures in turbulent boundary layers are highly entangled and interactions are frequent.

\begin{figure}[!t]
 \centering
    \includegraphics[width=0.99\linewidth]{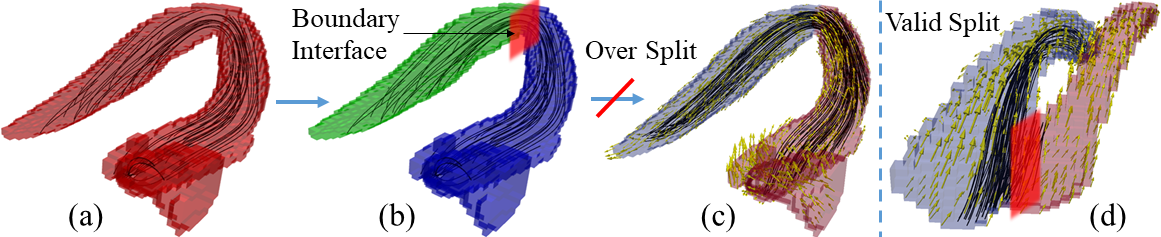}
    \caption{\revise{This figure summarizes the method proposed in~\cite{zafar2024topological} for addressing over-segmentation. (a) shows a hairpin vortex region. (b) shows two split segments (green and blue) obtained via minimal merge tree-based segmentation and layering, with their boundary highlighted (red square). (c) Vortex lines seeded from the boundary interface cover approximately equal areas in both segments, indicating an invalid split that is therefore reversed, leaving the hairpin vortex intact. (d) In contrast, a valid split between two streamwise vortices is shown, where the vortex lines predominantly cover only one segment.}}
 \label{fig:3_3A}
\end{figure}

Recently, Zafar et al.~\cite{adeelhairpin2023} introduced a method to aid the identification of hairpin vortices from turbulent flow with a relatively low Reynolds number~\cite{adeelhairpin2023}.
\revise{In their framework, 
vortices are first extracted using a $\lambda_2$-based thresholding approach, where the 90th bin (out of 100) of the $\lambda_2$ histogram is used to identify vortical regions. Entangled vortices are then separated by progressively extracting isosurfaces at selected $\lambda_2$ values, producing a hierarchical segmentation tree in which leaf nodes represent the final split vortices. The extent of splitting is controlled by the Vortex Size Factor (VSF). Skeletons are computed for the leaf vortices, and those with high curvature and positive $\omega_y'$ are selected as candidate hairpin vortices. For each candidate, a feature-based vortex profile is constructed and clustered to group similar structures. The results are explored using an interactive visualization system that links 3D views with hierarchical and feature-based representations, enabling the identification of hairpin vortices across clusters.}

\revise{Despite its effectiveness, the above method~\cite{adeelhairpin2023} suffers from several limitations that are closely related to the under- and over-segmentation issues discussed earlier. First, segmentation based on a fixed list of $\lambda_2$ values can cause abrupt splitting, producing many small segments and leading to over-segmentation of hairpin vortices. Attempting to recover intact structures by selecting parent nodes in the Tree View often reconnects unrelated vortices, resulting in under-segmentation. Second, when hairpin vortices are fragmented, the resulting skeletons may fail the curvature and $\omega_y' > 0$ criteria, leading to a high rate of false negatives. Third, the candidate selection stage produces many false positives, which propagate through clustering and tree exploration and significantly increase analysis effort. Finally, the method relies on a user-driven feedback loop involving feature selection, clustering, parameter tuning, and repeated visual inspection, making the extraction process time-consuming and subjective.}

\revise{Later, Zafar et al. \cite{zafar2024topological} proposed a merge tree-based vortex separation method to address over-segmentation by exploiting the property that each merge tree level produces at most two branches, mitigating abrupt segmentation as shown in \cref{fig:3_1A}((c) and (g)). 
The method begins by constructing a minimal merge tree for each region by selecting the most persistent~\cite{edelsbrunner2022computational} maximum-minimum and saddle-minimum pairs while removing other critical points via topological simplification, restricting segmentation to three regions as shown in \cref{fig:3_1A}((b) and (c)). A layering step then expands the leaf segments (blue and red segments in \cref{fig:3_1A}(b)) into the surrounding parent segment (green segment in \cref{fig:3_1A}(b)) until the parent segment disappears, leaving a linear split into only two segments (\cref{fig:3_1A}(e)). The split segments form a boundary interface (illustrated for a hairpin vortex in \cref{fig:3_3A}(b)) from which vortex lines are seeded (\cref{fig:3_3A}(c)). A vortex line-based statistic computes the ratio of coverage in the two segments; if below a threshold, the split is rejected (\cref{fig:3_3A}(c)), otherwise accepted (\cref{fig:3_3A}(d)). This process repeats until all minimum-saddle pairs are exhausted.}

\begin{figure}[!t]
 \centering
    \includegraphics[width=0.99\linewidth]{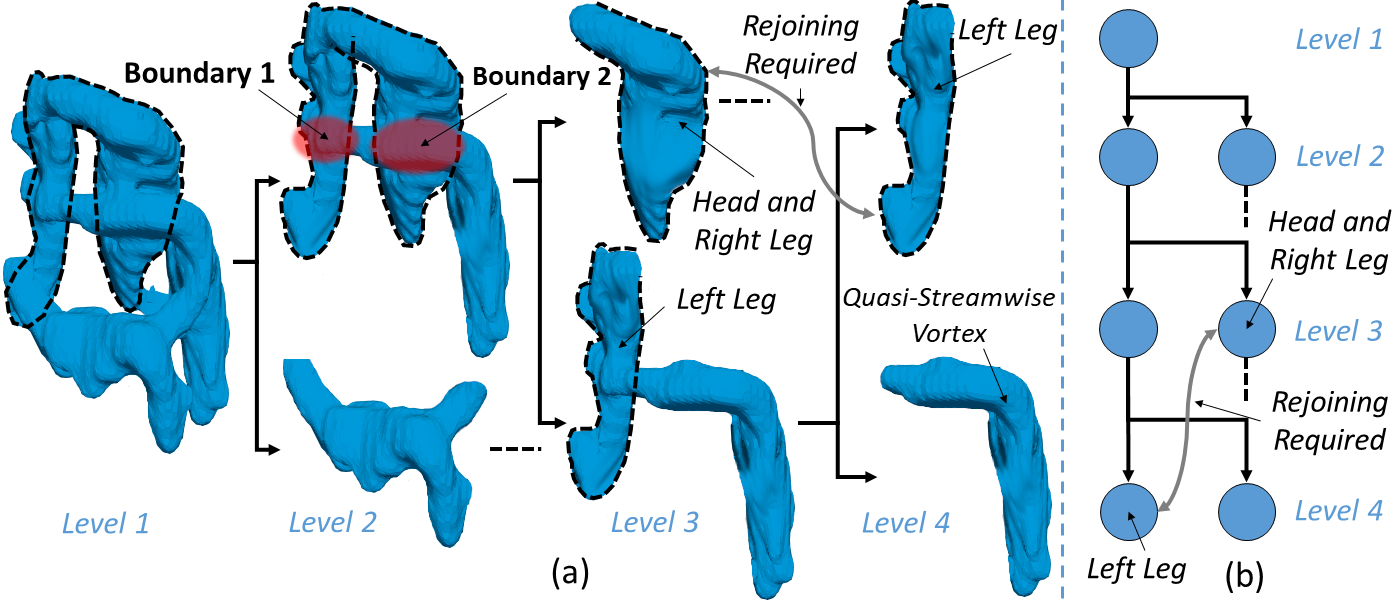}
    \caption{\revise{This figure shows the failure of the method in \cite{zafar2024topological} to separate a hairpin vortex from a vortical region. (a) A hairpin vortex is highlighted at level 1, where it is attached to other vortices at multiple boundary interfaces. By level~3, the left leg of the hairpin vortex detaches from the rest of the structure causing over-segmentation. (b) shows a simplified tree view of the split hierarchy of the vortical region.}}
 \label{fig:3_4A}
\end{figure}

\revise{The method assumes a single boundary interface between split segments (\cref{fig:3_3A}(b)), which is often violated in highly entangled vortices connected at multiple interfaces, causing the vortices to over-split (\cref{fig:3_4A}(a)). Here, the left leg of a hairpin vortex is prematurely split from the main body at level~3 because its connection to a neighboring streamwise vortex is stronger (\cref{fig:3_4A}(a), level 3 bottom). Consequently, the hairpin vortex may split before full separation, requiring identification and rejoining of segments across multiple levels of the split tree (\cref{fig:3_4A}(b)) to reconstruct the complete vortex. This demonstrates that top-down hierarchical splitting alone is insufficient to resolve under- and over-segmentation. Moreover, the method \cite{zafar2024topological} is not tailored for hairpin vortices, focusing instead on general vortex separation in relatively low Reynolds number turbulent flows, and it incurs high computational cost due to repeated minimal merge tree construction, segmentation, layering, and recursive split validation.}

\begin{figure*}[!t]
 \centering
    \includegraphics[width=0.99\linewidth]{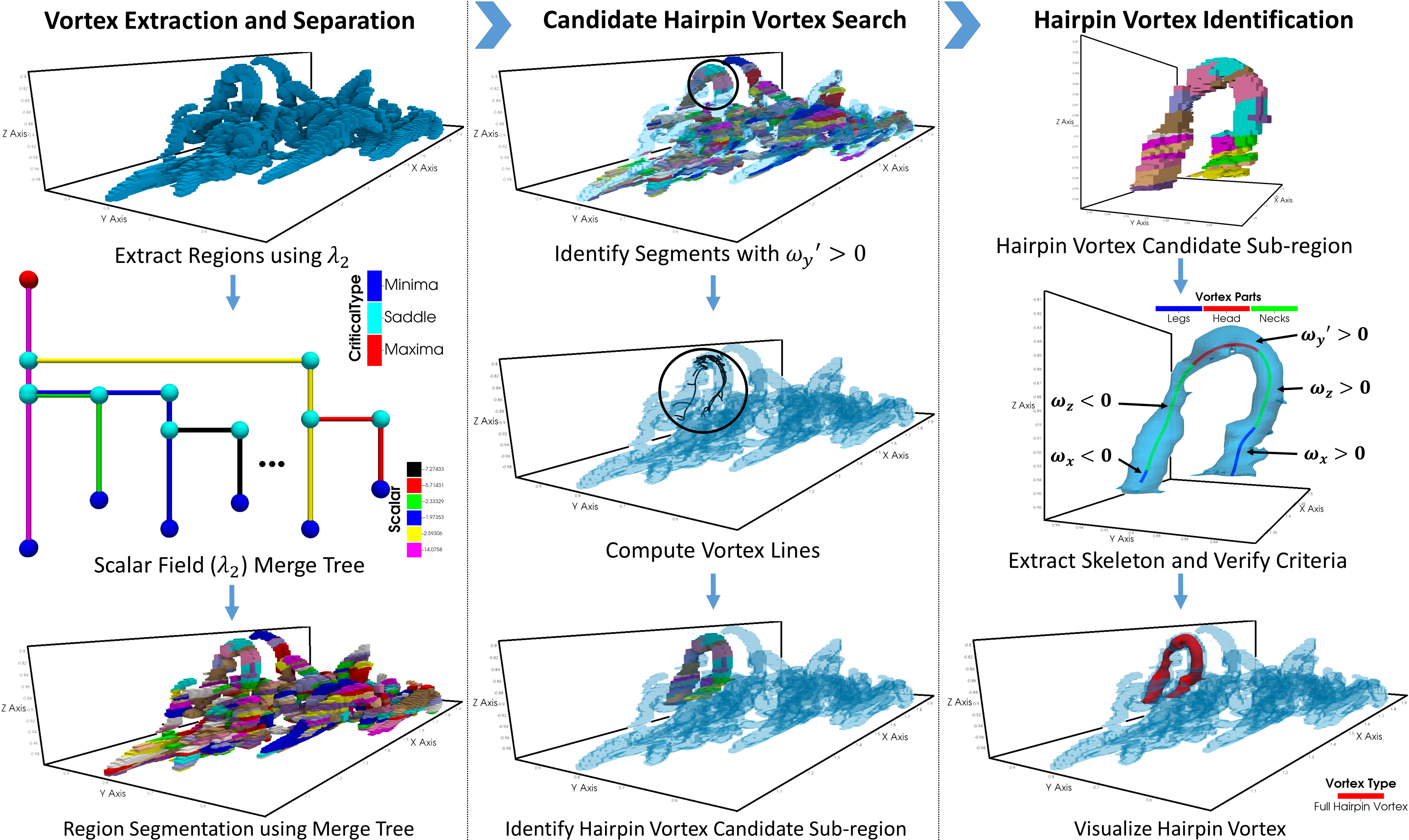}
    \caption{The pipeline of our framework for the extraction and visualization of hairpin vortices.}
 \label{fig:3_1}
\end{figure*}

\revise{The two methods~\cite{adeelhairpin2023, zafar2024topological} inadequately address under- and over-segmentation, partly because segmentation and over-segmentation avoidance are combined into a single top-down process. While Zafar et al.~\cite{adeelhairpin2023} used a Tree View to manually handle over-segmentation, which in turn may cause under-segmentation, Zafar et al.~\cite{zafar2024topological} addressed only simple cases. Both suffer from high false negative rates due to under- and over-segmentation (\cref{sec:results}). In contrast, in this study we propose a top-down segmentation followed by a bottom-up rejoining step that automatically identifies and reunites split hairpin vortex segments (\cref{fig:4_1A}(b)), enabling the extraction of intact hairpin structures. Our new method offers an automated pipeline for hairpin vortices extraction.}

%% file: content/method.tex
\section{Automated Hairpin Vortex Extraction}
\label{sec:method}

\begin{figure*}[!t]
 \centering
    \includegraphics[width=0.95\linewidth]{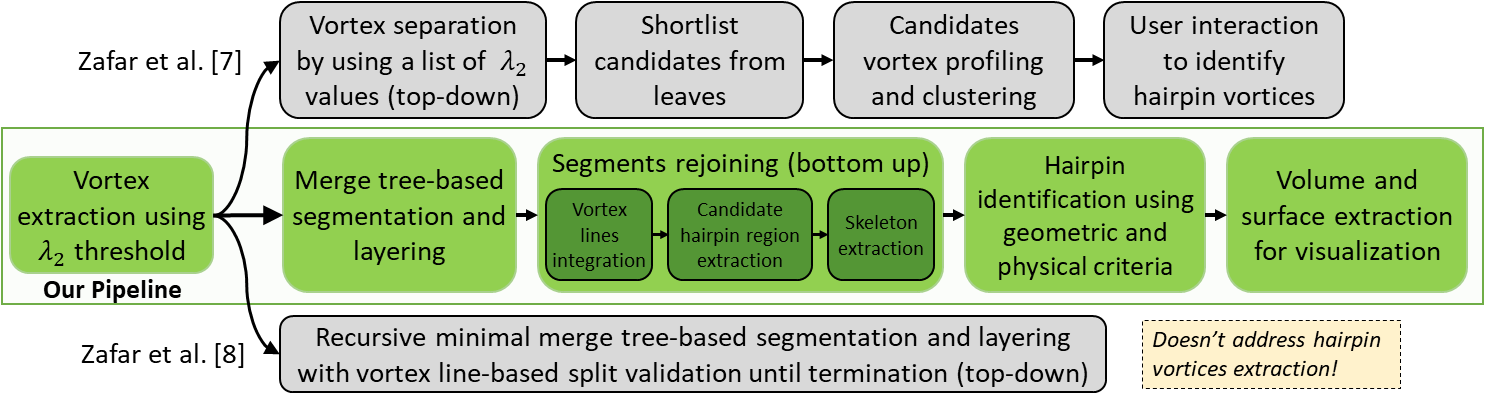}
    \caption{\revise{This figure presents a comparison chart of the steps in our pipeline with those of Zafar et al.~\cite{adeelhairpin2023} and Zafar et al.~\cite{zafar2024topological}.}}
 \label{fig:3_5A}
\end{figure*}

This section details our framework for the automated extraction of hairpin vortices, as outlined in \cref{sec:intro} and illustrated in \cref{fig:3_1}. The framework begins by identifying vortical regions and dividing them into smaller segments to ensure sufficient separation of different vortex structures (\cref{subsec:segmentation}). Next, it rejoins specific segments into candidate hairpin vortex regions, which are then verified and refined based on the geometric and physical criteria of hairpin vortices (\cref{subsec:hairpincriteria} and \cref{subsec:hairpinIdentification}). Finally, a surface enclosing the underlying hairpin vortex volume is extracted to enable effective visualization (\cref{subsec:hairpinVisualization}). Figure \ref{fig:3_5A} compares our pipeline with the previous methods. In the following subsections, we provide more details on each of these key steps.

\begin{figure}[!t]
 \centering
    \includegraphics[width=0.99\linewidth]{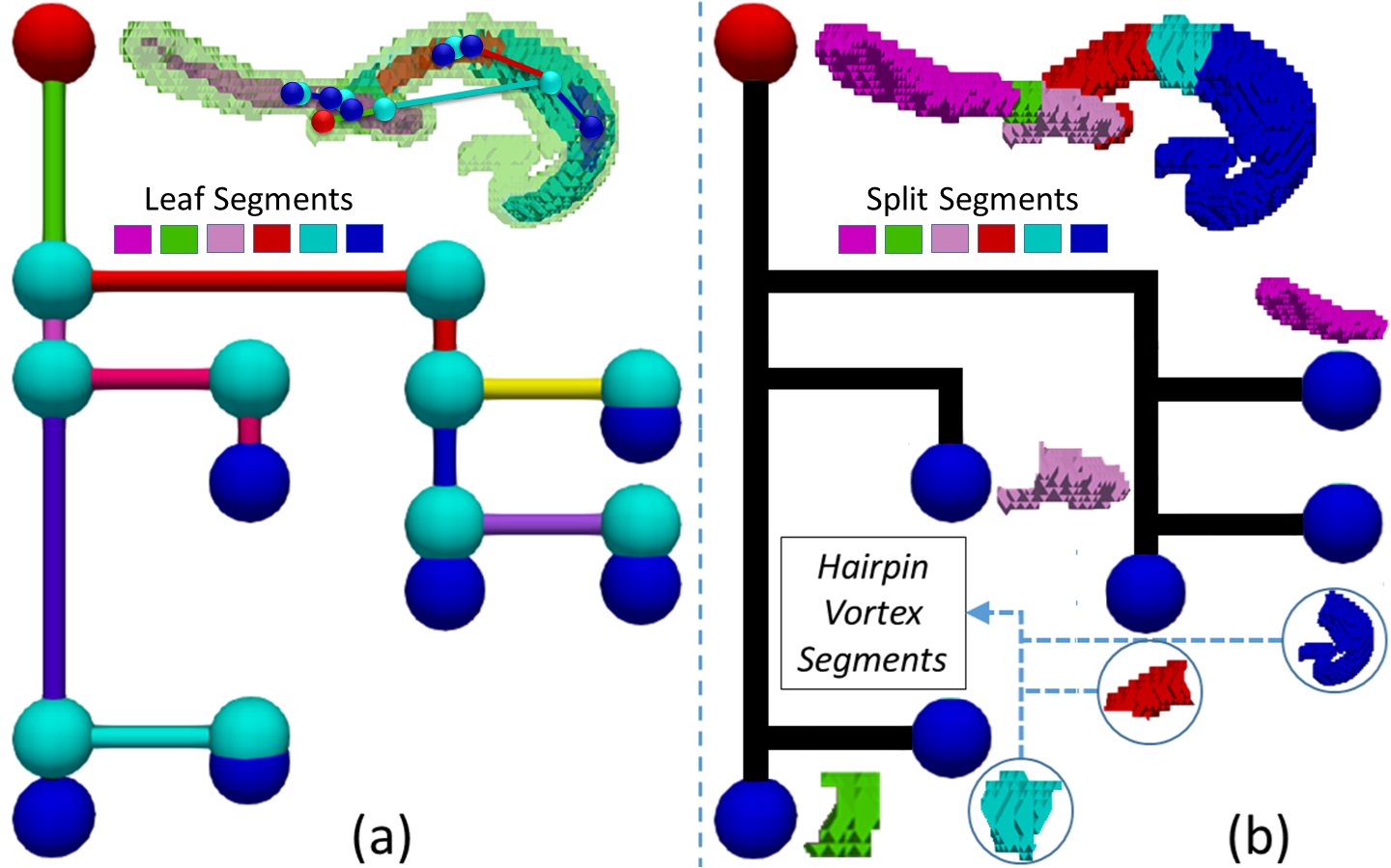}
    \caption{\revise{This figure illustrates the single-pass layering process applied to the entire merge tree-based segmentation. (a) shows the initial segmentation of the vortical region (b) presents the final segmentation produced after layering. The leaf segments corresponding to the six saddle–minima edge pairs in (a) are iteratively expanded into their respective parent segments (saddle–saddle pairs) using the layering process until all parent segments disappear, thereby converting the hierarchical merge tree-based segmentation into a linear segmentation. The hairpin vortex split segments are highlighted in (b).}}
 \label{fig:4_1A}
\end{figure}

\subsection{Vortex Extraction and Segmentation}
\label{subsec:segmentation}
Similar to the above methods, we extract vortical regions using $\lambda_2$ as the scalar criterion. For segmentation, we adopt the merge tree-based segmentation and layering process from~\cite{zafar2024topological}, with slight modifications. In the original method, the vortical region is recursively split into exactly two segments at each step using the minimal merge tree (\cref{fig:3_1A}((a)-(e))). In contrast, our approach extracts the entire merge tree and performs the merge tree-based segmentation only once (\cref{fig:4_1A}(a)). As in~\cite{zafar2024topological}, from the segmentation, we identify the leaf segments corresponding to saddle-minimum pairs in the merge tree. These leaf segments are then iteratively expanded through the layering process until convergence (\cref{fig:4_1A}(b)). This procedure is equivalent to converting the full hierarchical merge tree-based segmentation into a linear segmentation, without explicitly building a minimal merge tree and splitting the region into exactly two new segments repetitively. As a result, the computational cost is significantly reduced while producing a large number of fine-grained segments. At this stage, our objective is to obtain the finest possible level of splitting (\cref{fig:4_1A}(b)) without concern for over-segmentation. Subsequently, we introduce a bottom-up rejoining step that identifies and merges segments belonging to hairpin vortices (\cref{fig:4_1A}(b)), enabling efficient extraction of these structures (\cref{subsec:hairpinIdentification}). This decoupling of segmentation and rejoining provides added flexibility. We first mitigate under-segmentation through intentional over-splitting and then address over-segmentation by automatically rejoining overly split regions.

\begin{figure}[!t]
 \centering
     \includegraphics[width=0.99\linewidth]{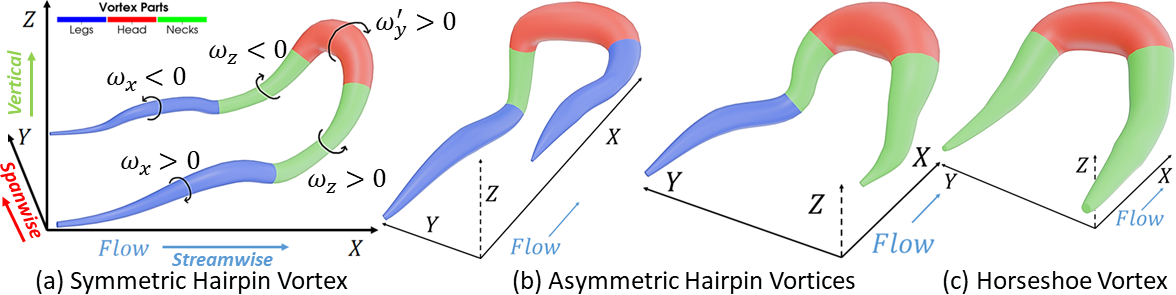}
     \caption{This figure illustrates the idealized shape of hairpin vortices. (a) shows the symmetric “hairpin” shape. (b) depicts asymmetric cases, also referred to as partial hairpin vortices. (c) represents the “horseshoe” or “$\Omega$-shaped” vortex.}
 \label{fig:3_3}
\end{figure}

\subsection{Hairpin Vortex Characteristics}
\label{subsec:hairpincriteria}
Ideally, a symmetric hairpin vortex resembles a ``hairpin,'' with a well-defined head, necks, and trailing legs that are evenly distributed on either side, as shown in \cref{fig:3_3}. In boundary layers, the head of the hairpin vortex is oriented in the spanwise direction, the necks in the quasi-vertical direction, and the legs in the quasi-streamwise direction upstream of the head. If $x$, $y$, and $z$ represent the streamwise, spanwise, and vertical directions, respectively, then ideally, a hairpin vortex fulfills the following positional and physical criteria.
\begin{itemize}[leftmargin=*]
  \setlength{\itemsep}{1pt}
  \setlength{\parskip}{1pt}
    \item[i] The head of the hairpin vortex has a positive spanwise vorticity fluctuation,  i.e.\ $\omega_{y}' > 0$. 
    \item[ii] The head is located slightly above the necks and the legs. Additionally, the legs always trail behind the head in the upstream direction. If $(H_x, H_y, H_z)$, $(N_x, N_y, N_z)$, and $(L_x, L_y, L_z)$ represent the coordinates of the centers of the bounding boxes of the head, necks, and legs, respectively, then $H_x > N_x > L_x$ and $H_z > N_z > L_z$. That is, the head is ahead of the necks, and the necks are ahead of the legs in the downstream direction, likewise, the head is above the necks and the necks are above the legs in the vertical direction, as shown in \cref{fig:3_3}(a).
    \item[iii] Since there are two necks and two legs, one on each side of the head, they can be further classified into the left neck and leg, and the right neck and leg, as viewed when facing the downstream direction of the flow (i.e., \ $+x$-direction). 
    \item[iv] The fluid in the left leg and neck rotates anticlockwise, while the fluid in the right leg and neck rotates clockwise. Therefore, the left leg has $\omega_{x} < 0$, the right leg has $\omega_{x} > 0$, the left neck has $\omega_{z} < 0$, and the right neck has $\omega_{z} > 0$ (\cref{fig:3_3}(a)), in accordance with the right-hand rule of vorticity.
\end{itemize}

However, at high Reynolds numbers, the increased turbulence intensity and chaotic flow interactions often distort this symmetry, leading to the asymmetric shapes. The necks or legs may disappear (\cref{fig:3_3}((b,c))), and the legs can vary in length or orientation due to the influence of surrounding vortical structures and flow instabilities. We incorporate these characteristic shapes and features into the next step of our pipeline to automate the extraction of hairpin vortices.

\begin{figure}[!t]
 \centering
    \includegraphics[width=0.99\linewidth]{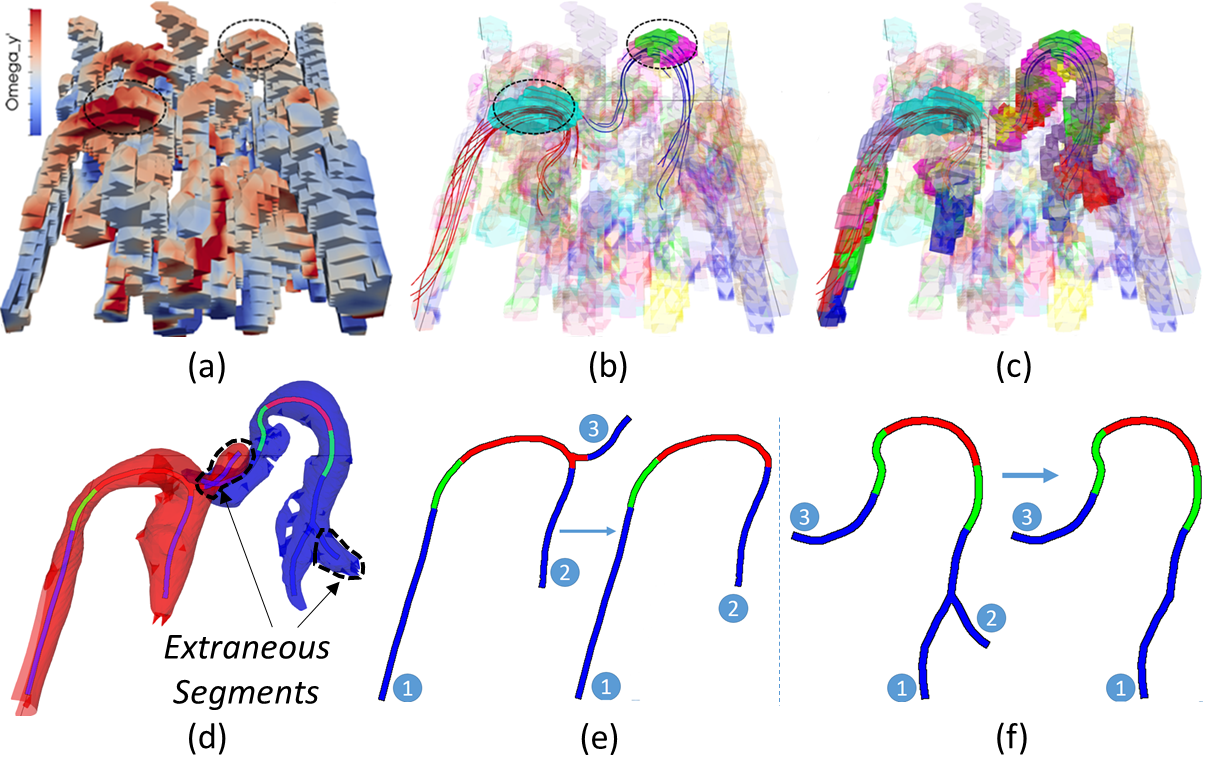} 
    \caption{This figure demonstrates the hairpin vortex identification process as explained in \cref{subsec:hairpinIdentification}. Areas with $\omega_y' > 0$ and higher $z$ elevation in the vortical region are highlighted in (a). (b) shows the segments with $\omega_y' > 0$ in the same areas highlighted in (a), along with the extracted vortex lines. (c) displays separate groups of segments overlapping with the vortex lines, shown with higher opacity. (d) presents the surfaces (red and blue) of the candidate hairpin vortex regions formed by joining each separate group of overlapping segments. (e) shows the skeleton of the red surface (left) and the selected hairpin sub-skeleton (right) fulfilling the criteria of a hairpin shape between the path of endpoints 1 and 2. (f) illustrates the same for the blue surface, where the sub-skeleton also fulfills the hairpin shape criteria.}
 \label{fig:3_4}
\end{figure}

\subsection{Hairpin Vortex Identification}
\label{subsec:hairpinIdentification}
Let $[z_{min}, z_{max}]$ denote the range of the vertical coordinate of the dataset, where the solid wall bounding the flow is located at $z_{min}$. The heads of hairpin vortices are therefore located at higher $z$-values. For Couette and channel flows, $z_{max}$ represents the height of the lower half of the domain. We first identify the segments corresponding to the heads of hairpin vortices, i.e., those split segments having $\omega_{y}' > 0$. Starting from $z_{max}$ and iterating toward $z_{min}$, we sequentially identify segments with $\omega_{y}' > 0$ and apply the following steps to determine whether a hairpin vortex exists in the vicinity.

\begin{figure}[!t]
 \centering
    \includegraphics[width=0.99\linewidth]{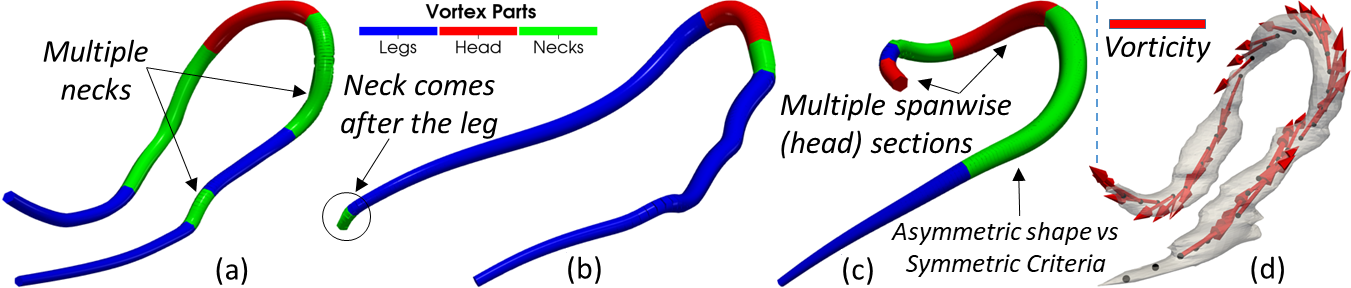} 
    \caption{This figure illustrates several practical cases of hairpin skeletons. (a) A hairpin skeleton with two neck sections. (b) A hairpin skeleton in which the leg section precedes the neck section. (c) A hairpin skeleton that fulfills the criteria of a symmetric hairpin vortex but corresponds to an asymmetric hairpin vortex. (d) shows the direction of the vorticity vectors along the skeleton.}
 \label{fig:4_3A}
\end{figure}

\begin{itemize}[leftmargin=*]
  \setlength{\itemsep}{1pt}
  \setlength{\parskip}{1pt}
    \item[1] For each segment with $\omega_{y}' > 0$~(\cref{fig:3_4}(a, b)), we take its points as seeds to compute vortex lines (also called vorticity lines) using the vorticity vector field $\vec{\omega}$. We use Runge-Kutta45~\cite{Fehlberg1969KlassischeRF} for its adaptive capability and integrate the vortex lines in both forward and backward directions~(\cref{fig:3_4}(b)).
    \item[2] Vortex lines tend to follow the shape of the vortex~\cite{Moin_Kim_1985}. Therefore, we identify the segments overlapping with the vortex lines and combine them to form a candidate hairpin region~(\cref{fig:3_4}(c)). \change{Here, \textit{overlap} is defined as a line point lying inside any cell of the segment.} This is equivalent to identifying the segments  from different scalar field levels and rejoining them~(\cref{fig:4_1A}(b)). The vortical structures are highly intertwined, and the extracted vortex lines may overlap with structures beyond the hairpin vortex. This attaches extraneous segments from nearby vortical structures~(\cref{fig:3_4}(d)), therefore, the candidate hairpin region needs to be further refined.
    \item[3] To do this, we extract skeleton using the mean-curvature skeletonization (MCS) method~\cite{taglia_sgp12} by extracting and smoothing the surface of the candidate hairpin region.
    \item[4] The resulting skeleton may contain multiple branches due to the extraneous segments~(\cref{fig:3_4}(d)). We need to find the sub-skeleton that fulfills the geometric and physical criteria of the hairpin vortex. Therefore, we traverse paths (\cref{fig:3_4}(e, f)) between all ends of the skeleton using the shortest geodesic distance and test which of these paths (sub-skeletons) fulfill one of the hairpin's shape and physical criteria from \cref{fig:3_3}.
    \item[5] For each sub-skeleton identified in Step-4, we further divide it into the head (spanwise), neck (vertical), and leg (streamwise) parts (\cref{fig:3_4}(e, f)). To achieve this, we calculate the dot product of each edge of the skeleton's polyline with $\hat{x}$, $\hat{y}$, and $\hat{z}$ unit vectors and use the following equations to decide to which part the edge belongs:
    \begin{equation}
        T = \textbf{argmax}((\hat{v}.\hat{x}), (\hat{v}.\hat{y}), (\hat{v}.\hat{z})) \quad and \quad |T| \geq 0.5
        \label{eq1}
    \end{equation}
    where $\hat{v}$ is the edge unit vector. From \cref{eq1}, $T = [0, 1, 2]$ corresponds to the leg, head, and neck, respectively.
    \item[6] After separating the sub-skeleton into the corresponding parts, we first check whether the sub-skeleton contains a head, necks, and legs. Moreover, each leg, neck, and head is individually checked against the corresponding vorticity criteria~(\cref{subsec:hairpincriteria}). If multiple sub-skeletons fulfill these criteria, we choose the one with the longest length.
    \item[7] We introduce a criterion called Consistency ($C$), a scalar measure that quantifies the alignment between the hairpin vortex skeleton and the local vorticity direction (\cref{fig:4_3A}(d)). It is defined as the average dot product between the skeleton’s edge unit vectors and the corresponding vorticity unit vectors, i.e., $C = \frac{\sum{(\hat{v} \cdot \hat{\omega})}}{n-1}$, where $n$ is the number of points in the skeleton. Ideally, for a valid hairpin vortex, $C \approx 1$. \change{However, in turbulent flows, extracted hairpin vortices are frequently distorted (\cref{sec:results}), resulting in an optimal value for $C$ between 0.8 and 0.85, as reported in \cref{tab:Table2}.}
\end{itemize}

\begin{figure}[!t]
 \centering
    \includegraphics[width=0.99\linewidth]{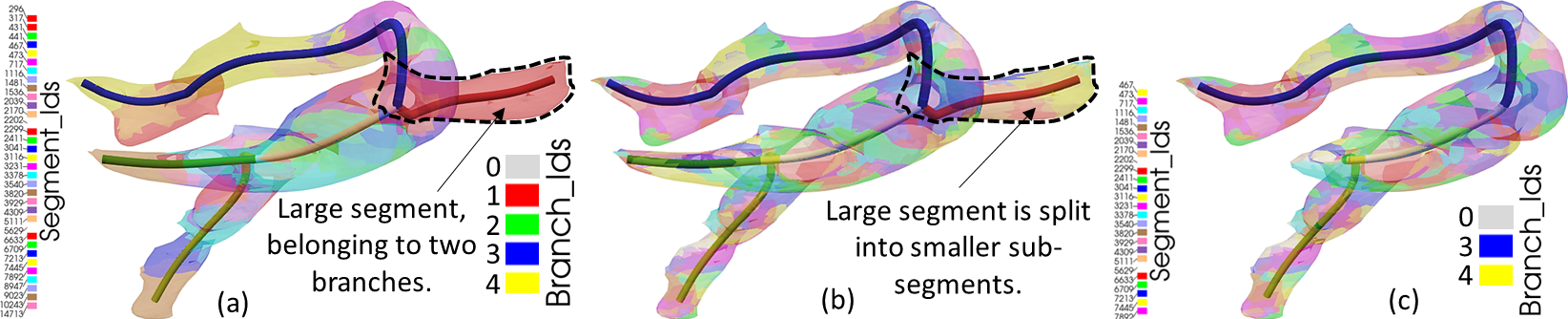} 
    \caption{This figure illustrates the hairpin vortex volume extraction process. (a) depicts the segments belonging to the candidate hairpin vortex region along with the skeleton and its individual branch IDs. (b) shows the finer sub-segments obtained after performing segmentation based on the $|\vec{\omega}|$ criterion. (c) shows the final sub-segments of the extracted hairpin vortex region and the branch IDs of the resulting hairpin vortex sub-skeleton. The segments associated with the removed branches have been deleted, and only the segments corresponding to the retained branches of the final hairpin vortex sub-skeleton are preserved. This process is performed on the volumetric data, however, a surface-based representation is shown here for clarity.}
 \label{fig:4_4A}
\end{figure}

The above steps are demonstrated in \cref{fig:3_4}. The hairpin characteristics described in \cref{subsec:hairpincriteria} are idealized. In practice, at Step-6, when the skeleton is divided into head, neck, and leg components, multiple candidates for each part may exist (\cref{fig:4_3A}(a)). In such cases, we select the longest streamwise sections on both sides as the legs, the longest vertical sections as the necks, and the spanwise section with the highest elevation as the head. Since a leg may precede the neck (\cref{fig:4_3A}(b)), we only require the existence of legs and necks and verify that they are located below the head. If this condition is satisfied, we then evaluate the vorticity-based criteria and proceed with the subsequent steps. Additionally, some skeletons may satisfy the criteria of a symmetric hairpin vortex but correspond to partial or asymmetric structures when the full vortex geometry is considered (\cref{fig:4_3A}(c)). This discrepancy arises from the sensitivity of the skeletonization process to small shape variations. Therefore, the mere presence of specific vortex body parts is insufficient to reliably distinguish between symmetric, asymmetric, or horseshoe subclasses. We leave the development of robust criteria for differentiating hairpin vortex subclasses for future work.

\subsection{Hairpin Vortex Visualization}
\label{subsec:hairpinVisualization}
\begin{figure}[!t]
 \centering
    \includegraphics[width=0.99\linewidth]{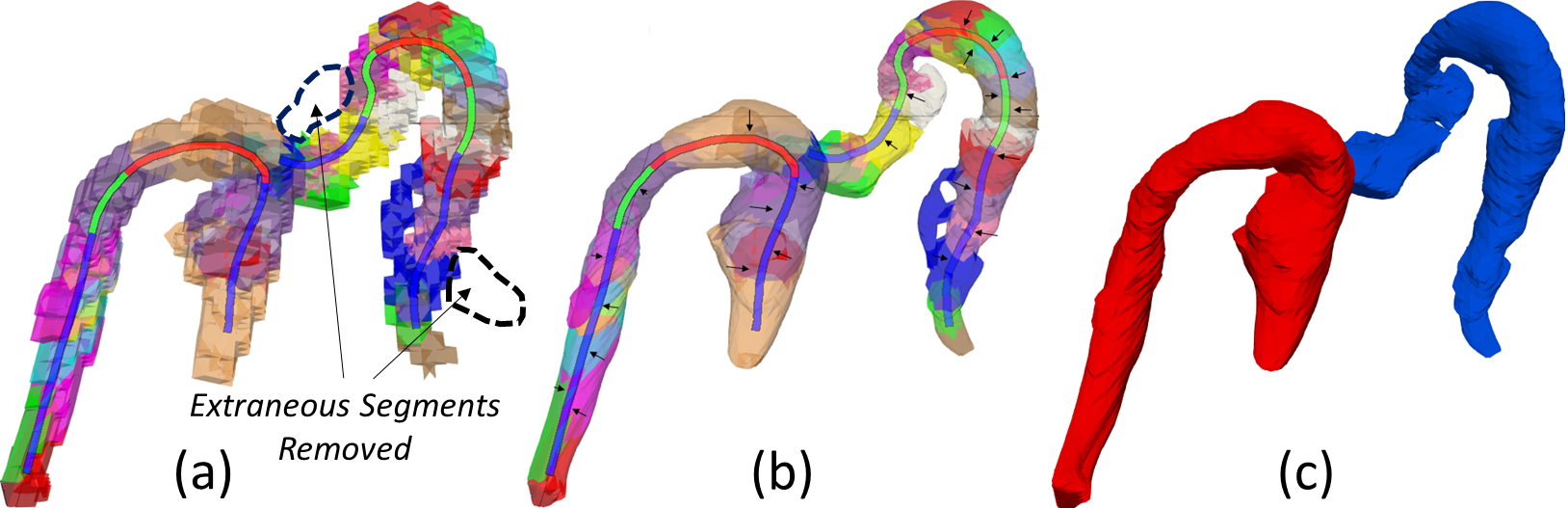} 
    \caption{This figure illustrates the hairpin vortex surface extraction process. (a) shows the final hairpin vortex volume. (b) shows the surface extracted from the final hairpin vortex volume. (c) shows the smoothed hairpin vortex surfaces rendered in separate colors.}
 \label{fig:3_4B}
\end{figure}

\revise{So far, we have ensured and verified the presence of a hairpin vortex-like shape in the candidate region by checking all hairpin vortex characteristics against the skeleton. However, we still need to extract the corresponding volume and visualize the resulting hairpin vortex. For this purpose, the skeleton extracted from the candidate hairpin region (Step-3) is divided into individual branches (\cref{fig:4_4A}(a)). Each branch is a polyline with no intersections and has a unique ID. For each branch, a list of nearest segments is computed. Once the final hairpin vortex sub-skeleton (\cref{fig:3_4}(e, f)) has been identified (Step-7), the segments corresponding to the identified hairpin sub-skeleton are merged to form the final hairpin vortex volume~(\cref{fig:3_4B}(b)). The segments corresponding to the skipped branches of the skeleton are also skipped, removing extraneous segments attached to the hairpin vortex~(\cref{fig:3_4B}(a)).}

In some cases, a segment may be too large and belong to multiple skeleton branches. For example, in \cref{fig:4_4A}(a), a large red segment is shared by both branch~1 and branch~3. When branch~1 is removed and branch~3 is retained in the final hairpin sub-skeleton, ambiguity arises as to whether the shared segment should be kept or discarded. Retaining it leads to under-segmentation, whereas removing it results in over-segmentation. To resolve this issue, we further subdivide the candidate hairpin region by applying merge tree–based segmentation and layering using magnitude of vorticity ($|\vec{\omega}|$) as the scalar criterion. We choose $|\vec{\omega}|$ because, compared to $\lambda_2$, it exhibits more local minima, producing finer segments. This refinement separates the portion of the large segment associated with the retained branch (e.g., branch~3) from that of the removed branch (e.g., branch~1), ensuring that only the appropriate sub-segments are preserved in the final volume.

\noindent \change{A practical question is why $|\vec{\omega}|$ is not employed as the initial base criterion.} First, when identifying segments that overlap with vortex lines to form the candidate hairpin region (Step 2), the likelihood increases that small segments belonging to the hairpin vortex may be missed (\cref{sec:limitations}). While this issue can also occur when using $\lambda_2$, it remains a primary limitation of our approach (\cref{sec:limitations}). However, since segments derived from the $\lambda_2$ scalar field are relatively large, the probability of missing relevant segments is significantly reduced. \change{As evidenced in \cref{tab:Table2}, the performance drops significantly when using $|\vec{\omega}|$ as the base criteria (e.g. $0.73$ vs. $0.24$ F1-score for TBL).} Therefore, finer segmentation using $|\vec{\omega}|$ is performed only after the candidate hairpin vortex region has been identified. Second, the larger number of segments generated by $|\vec{\omega}|$ produces many more candidates for vortex line and skeleton extraction, as illustrated in \cref{fig:4_4A}(b), which significantly increases computational costs. \change{As shown in \cref{tab:Table1}, the total number of segments increases nearly threefold, resulting in a significant increase in computational cost (e.g. $134$ vs $217$ minutes for TBL).} After this refinement process, the final hairpin vortex volume is obtained (\cref{fig:4_4A}(c)).

Isosurface-based visualization of vortices is one of the most widely adopted vortex visualization methods in the physics and visualization literature. However, since we have identified hairpin vortices by joining segments from different levels of the merge tree with varying $\lambda_2$ values (\cref{fig:4_1A}(b)), directly extracting the $\lambda_2$-based isosurface is not straightforward. Moreover, the $\lambda_2$-based isosurface may contain holes because we have separated the hairpin vortices from the surrounding overlapping structures and because of some missing segments. To overcome these challenges while retaining a surface representation for vortex visualization, we extract a pseudo-surface for visualizing the extracted hairpin vortex. We first extract the surface of the final hairpin vortex volume (\cref{fig:3_4B}(a,b)) and then apply Laplacian smoothing (\cref{fig:3_4B}(c)). Each surface is assigned a unique ID to distinguish individual hairpin vortices (\cref{fig:3_4B}(c)). These pseudo-surfaces form the closest upper bounds of the final hairpin vortex volume (\cref{fig:3_4B}(a)) and approximate isosurface-based visualization without holes.

%% file: content/results.tex
\section{Results and Evaluation}
\label{sec:results}
To demonstrate the effectiveness and robustness of our method, we present both quantitative and qualitative results on several turbulent flow datasets with varying Reynolds numbers and data sizes. Table \ref{tab:Table1} provides additional details and performance statistics of our method for these flow datasets.

\begin{table*}[!t]
\caption{\change{This table reports performance statistics for the flow datasets used in our experiments. \emph{Time} denotes the total runtime of the hairpin vortex search (in minutes), measured on a PC with an Intel Xeon(R) CPU E5-2640 without multiprocessing. The columns "No. Segments" and "Time (m)" compare the segment count and total computation time for the $\lambda_2$ and $|\vec{\omega}|$ criteria.}}
\centering
\resizebox{\textwidth}{!}{%
\begin{tabular}{|c|c|c|c|c|c|c|c|c|c|c|c|c|c|c|c|c|}
 \hline
 \multicolumn{1}{|c|}{Dataset} & 
 \multicolumn{1}{c|}{$Re_\tau$} & 
 \multicolumn{1}{c|}{Grid Size} & 
 \multicolumn{1}{c|}{$\lambda_2$} & 
 \multicolumn{3}{c|}{Segmentation time (m)} & 
 \multicolumn{2}{c|}{No. Segments} & 
 \multicolumn{1}{c|}{Lines} & 
 \multicolumn{1}{c|}{Segments} & 
 \multicolumn{1}{c|}{Surface} & 
 \multicolumn{1}{c|}{Skeleton} & 
 \multicolumn{1}{c|}{Skeleton} & 
 \multicolumn{1}{c|}{Post} & 
 \multicolumn{2}{c|}{Time (m)} \\
 \cline{5-9} \cline{16-17}
 & & & value &
 A~\cite{adeelhairpin2023} &
 B~\cite{zafar2024topological} &
 Ours &
 $\lambda_2$ & $|\vec{\omega}|$ &
 integ. & rejoining & extract. & extract. & search. & process & $\lambda_2$ & $|\vec{\omega}|$ \\
 \hline
 Couette & 180 & 384$\times$384$\times$193 & -0.0173 & 8 & 28 & \textbf{7} & \textbf{1215} & 14822 & 0.6 & 0.5 & 1 & 8 & 1 & 2 & \textbf{20} & 27 \\
 Channel & 1000 & 256$\times$192$\times$256 & -13.395 & 11 & 200 & \textbf{9} & \textbf{9289} & 24589 & 0.9 & 0.8 & 8 & 40 & 5 & 3 & \textbf{67} & 97 \\
 TBL & 1502 & 276$\times$251$\times$224 & -0.0272 & 41 & 388 & \textbf{12} & \textbf{8441} & 24385 & 1.2 & 1.1 & 12 & 72 & 32 & 4 & \textbf{134} & 217 \\
 \hline
\end{tabular}%
}
\label{tab:Table1}
\end{table*}

\begin{table}[!t]
\caption{\change{Statistical performance comparison of our method (C), Zafar et al.~\cite{adeelhairpin2023} (A), and Zafar et al.~\cite{zafar2024topological} (B) using reference hairpin vortices. This table integrates the sensitivity analysis of the parameter $C$ for the channel and TBL flows. The optimal configurations for our method ($C=0.8$ for channel, $C=0.85$ for TBL) are highlighted in bold. The results reported for the Couette flow do not represent the true statistics of the dataset, as the sample size is relatively small. For method A, the best VSF values were 3.5, 1.0, and 3.0 for the respective datasets. The average ratios of extra and missing points relative to the reference hairpin vortices are: Couette $(0.16, 0.04)$, channel $(0.25, 0.11)$, and TBL $(0.27, 0.09)$. The F1-score using ($|\vec{\omega}|$) as an alternative base criterion is also appended to demonstrate the superior performance of $\lambda_2$.}}
\label{tab:Table2}
\centering
\resizebox{\linewidth}{!}{%
\begin{tabular}{|l|ccccccc|}
\hline
\textbf{Method / Parameter} & \textbf{TP} & \textbf{FP} & \textbf{FN} & \textbf{Avg(IoU)} & \textbf{Precision} & \textbf{Recall} & \textbf{F1} \\
\hline
\hline
\multicolumn{8}{|c|}{\textbf{Couette (Total = 6)}} \\
\hline
Method A & 5 & 0 & 1 & 0.68 & 1.00 & 0.83 & 0.90 \\
Method B & 6 & 2 & 0 & 0.74 & 0.75 & 1.00 & 0.86 \\
Method C ($C=0.8$) & 6 & 2 & 0 & \textbf{0.85} & 0.75 & 1.00 & 0.86 \\
\hline
\multicolumn{8}{|c|}{\textbf{Channel (Total = 74)}} \\
\hline
Method A & 24 & 17 & 50 & 0.56 & 0.59 & 0.32 & 0.42 \\
Method B & 28 & 16 & 46 & 0.58 & 0.64 & 0.38 & 0.48 \\
Method C ($C=0.7$) & 60 & 30 & 14 & 0.72 & 0.67 & 0.81 & 0.73 \\
Method C ($C=0.75$) & 60 & 28 & 14 & 0.72 & 0.68 & 0.81 & 0.74 \\
\textbf{Method C ($C=0.8$)} & \textbf{63} & \textbf{24} & \textbf{11} & \textbf{0.71} & \textbf{0.72} & \textbf{0.85} & \textbf{0.78} \\
Method C ($C=0.85$) & 55 & 26 & 19 & 0.72 & 0.68 & 0.74 & 0.71 \\
Method C ($C=0.9$) & 53 & 19 & 21 & 0.71 & 0.74 & 0.72 & 0.73 \\
Method C ($|\vec{\omega}|$ base) & 22 & 53 & 52 & 0.57 & 0.29 & 0.30 & 0.30 \\
\hline
\multicolumn{8}{|c|}{\textbf{TBL (Total = 58)}} \\
\hline
Method A & 10 & 13 & 48 & 0.60 & 0.43 & 0.17 & 0.25 \\
Method B & 12 & 9 & 46 & 0.59 & 0.57 & 0.21 & 0.30 \\
Method C ($C=0.7$) & 47 & 41 & 11 & 0.71 & 0.53 & 0.81 & 0.64 \\
Method C ($C=0.75$) & 48 & 39 & 10 & 0.72 & 0.55 & 0.83 & 0.66 \\
Method C ($C=0.8$) & 51 & 34 & 7 & 0.72 & 0.60 & 0.88 & 0.71 \\
\textbf{Method C ($C=0.85$)}& \textbf{49} & \textbf{27} & \textbf{9} & \textbf{0.72} & \textbf{0.64} & \textbf{0.84} & \textbf{0.73} \\
Method C ($C=0.9$) & 43 & 25 & 15 & 0.72 & 0.63 & 0.74 & 0.68 \\
Method C ($|\vec{\omega}|$ base) & 14 & 43 & 44 & 0.57 & 0.25 & 0.24 & 0.24 \\
\hline
\end{tabular}%
}
\end{table}

\begin{figure}[!t]
    \centering
    \includegraphics[width=0.99\linewidth]{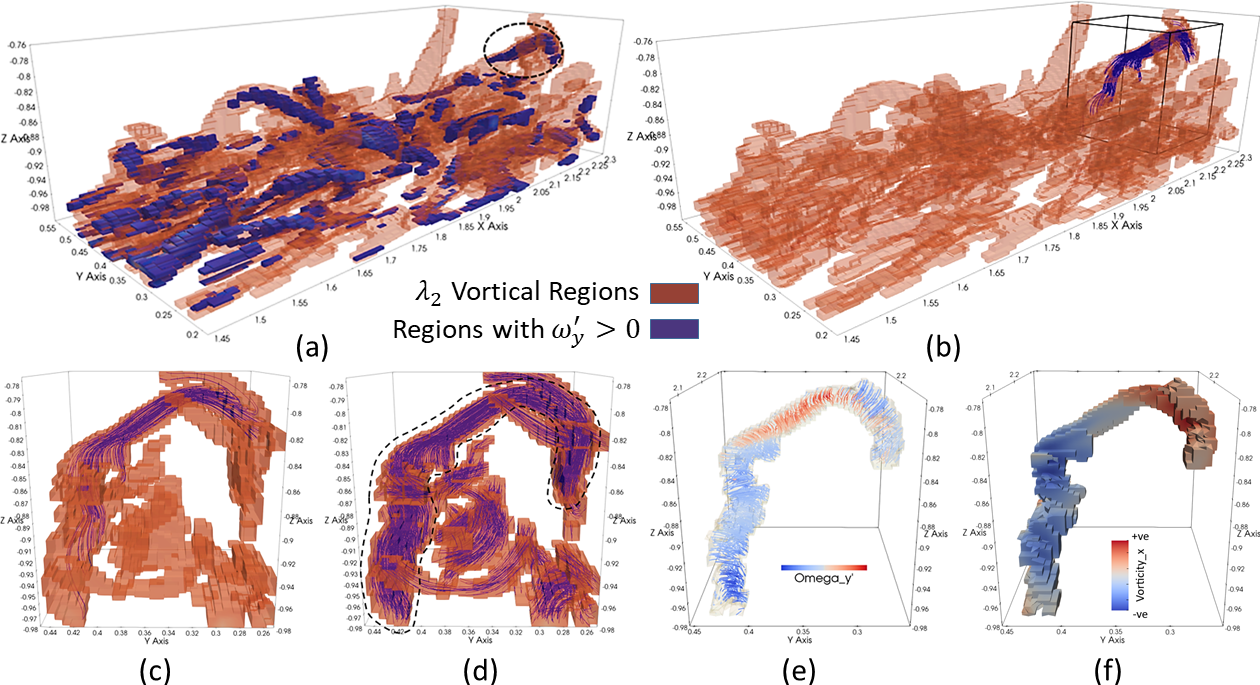}
    \caption{This figure illustrates the reference hairpin vortex extraction process. (a) A general vortical region (brown) extracted using $\lambda_2$, along with the regions where $\omega_y' > 0$ (blue) within the vortical region. (b) Vortex lines extracted using the highlighted region with $\omega_y' > 0$ in (a) as seeds. (c) The cut-off vortical regions confined within the bounds of the vortex lines from (b). (d) Vortex lines extracted using uniform seeding within the cut-off vortical region. (e) The separated hairpin vortex, showing swirling motion depicted by streamlines, with color indicating the distribution of $\omega_y'$. This segmented hairpin vortex serves as a reference for evaluation.}
    \label{figr:4_1}
\end{figure}

\subsection{Reference Hairpin Vortices Extraction}
\label{subsec:references}
For evaluation purposes, we extracted reference hairpin vortices from the datasets using the manual segmentation procedure illustrated in \cref{figr:4_1}. Within the vortical regions identified using $\lambda_2$, sub-regions with $\omega_y' > 0$ were first selected as potential candidates for hairpin vortex heads. Vortex lines were then generated by seeding from each head sub-region and visually examined to determine whether a hairpin vortex was present in the surrounding region. When a hairpin structure was confirmed, the vortical region was truncated using the extracted vortex lines as boundaries, thereby confining the region of interest. To capture internal coherent structures, additional vortex lines were generated through uniform seeding within the truncated regions. Finally, individual hairpin vortices were isolated by manually removing segments that did not belong to the vortex, based on visual analysis of which vortex-line sets formed a coherent hairpin pattern, as shown in \cref{figr:4_1}(d).

For each reference vortex, we apply a multi-criteria validation process to confirm that the structure represents a true hairpin vortex. Specifically, we examine: (1) coherent swirling streamline patterns (\cref{figr:4_1}(e)), (2) an organized arrangement of vortex lines forming a consistent hairpin topology (\cref{figr:4_1}(d)), (3) positive spanwise vorticity
fluctuation ($\omega_y' > 0$) in the head region (\cref{figr:4_1}(e)), (4) counter-rotating motions on the left ($\omega_x < 0$) and right branches ($\omega_x > 0$) (\cref{figr:4_1}(f)), and (5) an overall geometry resembling the canonical hairpin shape (\cref{fig:3_3}). Together, these criteria ensure that the segmented structure is a hairpin vortex rather than an artifact or partial fragment. This validation strategy is consistent with classical studies on hairpin vortices \cite{Head_Bandyopadhyay_1981, Moin_Kim_1985, robinson1991coherent, smith1991dynamics, perry1995wall}. We refer to these manually segmented structures as “reference” hairpin vortices because they are derived using the $\lambda_2$-based vortex definition. Consequently, they are not directly applicable if alternative criteria are used. Nevertheless, they provide a reliable foundation for both quantitative and qualitative comparisons with our method, prior approaches, and future methods that adopt the same $\lambda_2$-based definition.

Several steps in our pipeline automate the manual segmentation process. In Step-1 of \cref{subsec:hairpinIdentification}, segments with $\omega_y' > 0$ are used as seeds for vortex line extraction. Rather than directly using the $\omega_y' > 0$ regions shown in \cref{figr:4_1}(a), we operate on finer-grained segments, which are smaller than the corresponding regions. This increased granularity provides multiple opportunities to detect hairpin vortices, improving the likelihood of successful identification. In Step-2 of \cref{subsec:hairpinIdentification}, segments intersecting vortex lines are merged to form candidate hairpin sub-regions, mirroring the manual region-pruning step. The resulting collection of vortex lines forms a coherent hairpin pattern, and the subsequent skeleton extraction aggregates these lines into a 1D representation that enables automated shape analysis.

\subsection{Evaluation Metrics}
Based on the extracted reference hairpin vortices, we report the $F1$ score for each flow dataset in \cref{tab:Table2}, where

{\footnotesize
$\text{F1} = 2 \cdot \frac{\text{Precision} \cdot \text{Recall}}{\text{Precision} + \text{Recall}},\;
\text{Precision} = \frac{TP}{TP + FP},\;
\text{Recall} = \frac{TP}{TP + FN}$
}

A detection is considered a true positive (TP) if its intersection-over-union ($IoU$) with a reference hairpin vortex exceeds $0.5$. If a reference does not have any corresponding detection with $IoU > 0.5$, it is counted as a false negative (FN). Conversely, a detection that does not overlap with any reference hairpin vortex is counted as a false positive (FP). 

In previous methods, \cite{adeelhairpin2023} relied on user feedback to visually shortlist true hairpin vortices from a large pool of candidates, while \cite{zafar2024topological} does not target hairpin vortices. For fair comparison, we iterate over all candidates, i.e., all structures corresponding to the tree nodes in~\cite{adeelhairpin2023} and the final separated vortices in~\cite{zafar2024topological}, extract their skeletons, and apply our own search process (\cref{subsec:hairpinIdentification}) to identify potential hairpin vortices. We then determine which ones have $IoU > 0.5$ relative to the reference hairpin vortices. Moreover, in~\cite{adeelhairpin2023}, \textit{VSF} controls the degree of splitting, with a recommended value of $3.5$ for the Couette flow dataset. Since other datasets were not targeted in their study, we test values in $[2.5, 3.0, 3.5, 4.0, 4.5]$ and report the best results in \cref{tab:Table2}.


\begin{figure}[!t]
    \centering
    \includegraphics[width=0.99\linewidth]{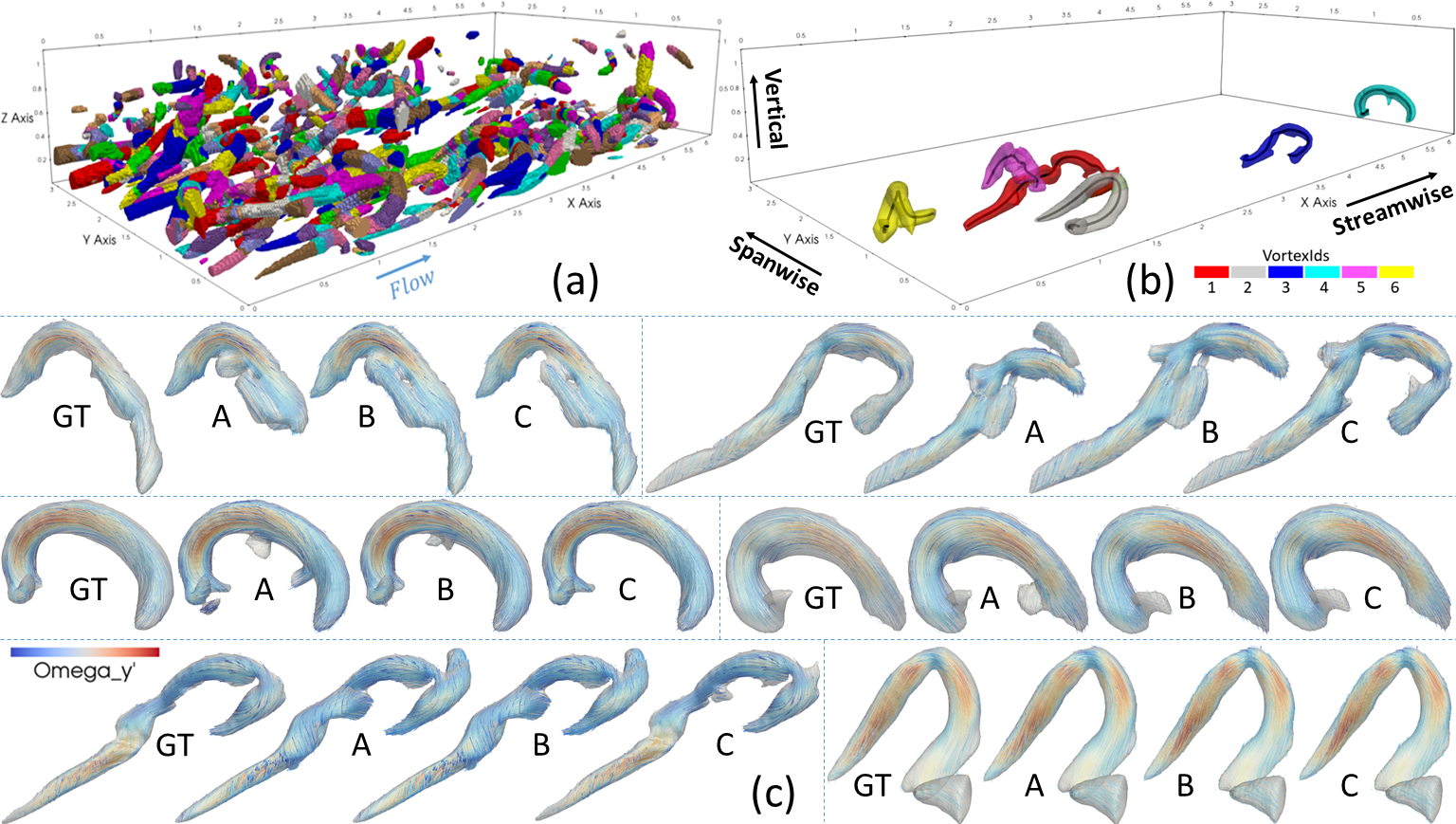}%
    \caption{\revise{This figure shows the extracted hairpin vortices from the bottom boundary layer of Couette flow from~\cite{li2019direct}. (a) shows the split segments of the general vortical regions obtained using our segmentation approach. (b) shows the extracted hairpin vortices. (c) presents a side-by-side comparison of each hairpin vortex between the references (GT), Zafar et al.~\cite{adeelhairpin2023} (A), Zafar et al.~\cite{zafar2024topological} (B), and ours (C). It can be seen that our method produces significantly cleaner and more intact structures compared to previous methods.}}
    \label{fig:5_1A}
\end{figure}

\subsection{Results on Couette Flow}
First, we apply our hairpin vortex identification method to analyze the direct numerical simulation (DNS) dataset of stress-driven turbulent Couette flow from~\cite{li2019direct}. In this Couette flow system, the fluid moves through the gap between two parallel boundaries. The bottom boundary is a no-slip impermeable flat surface, while the top boundary is a free-slip impermeable flat surface with a constant shear stress applied to it to drive the mean flow. Hereinafter, this dataset is referred to as the Couette flow dataset. This dataset is primarily used to provide a qualitative comparison between our approach and previous methods, as it contains only a small number of hairpin vortices. Our method achieves the highest value for $IoU$, which directly translates to improved extraction quality of the hairpin vortices as depicted in \cref{fig:5_1A}.

\begin{figure}[!t]
    \centering
    \includegraphics[width=0.99\linewidth]{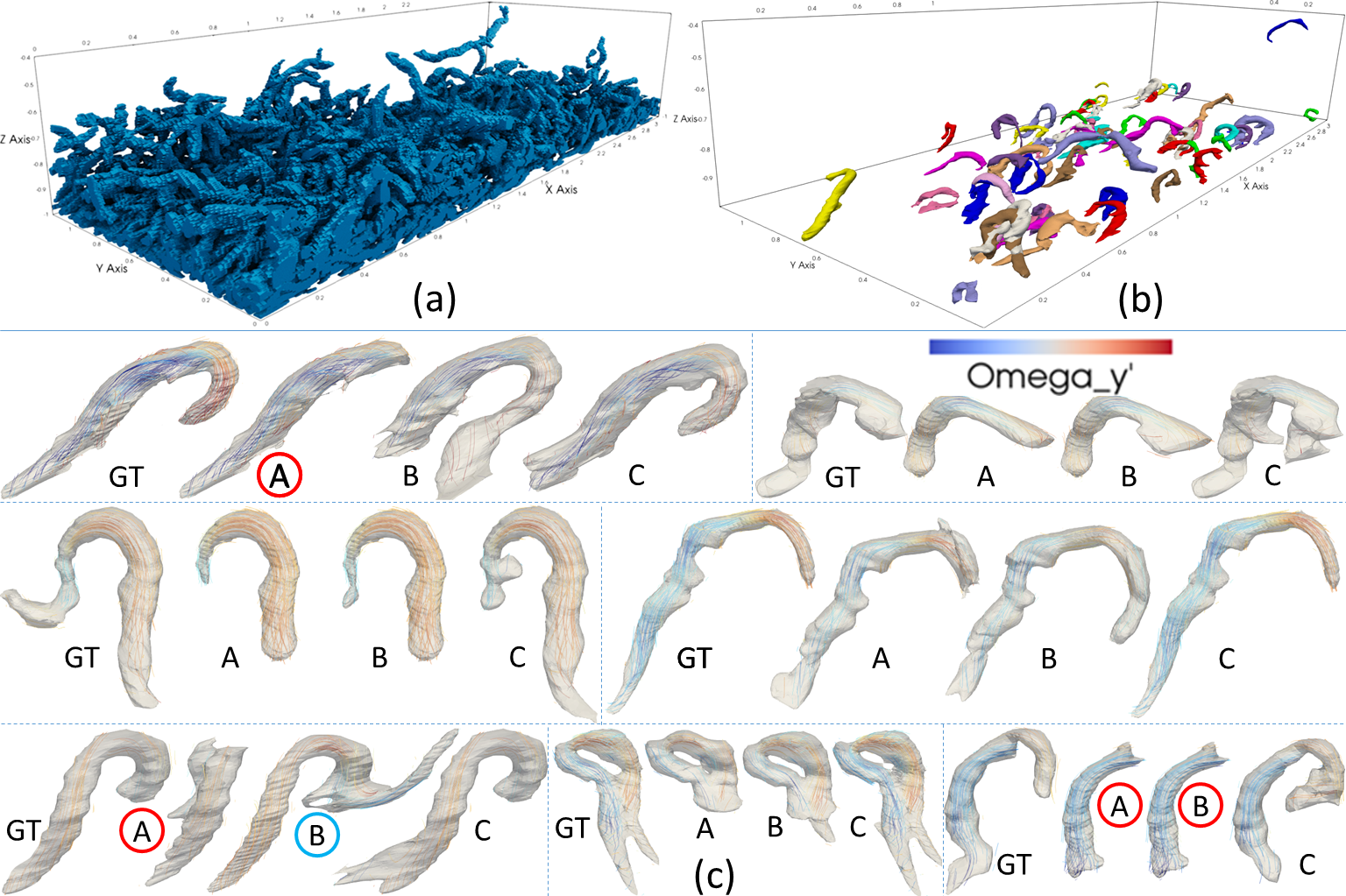}%
    \caption{\revise{This figure shows the extracted hairpin vortices from the channel flow~\cite{lee2013petascale}. (b) Individual vortices are depicted in distinct colors. (c) presents a side-by-side comparison of several hairpin vortices between the references (GT), Zafar et al.~\cite{adeelhairpin2023} (A), Zafar et al.~\cite{zafar2024topological} (B), and ours (C). Those highlighted with blue and red circles indicate severely under- and over-segmented cases of the previous methods, respectively. Color depicts the distribution of $\omega_y'$ within the structures, highlighting positive values around the head regions.}}
    \label{fig:5_2A}
\end{figure}

\begin{figure}[!t]
    \centering
    \includegraphics[width=0.99\linewidth]{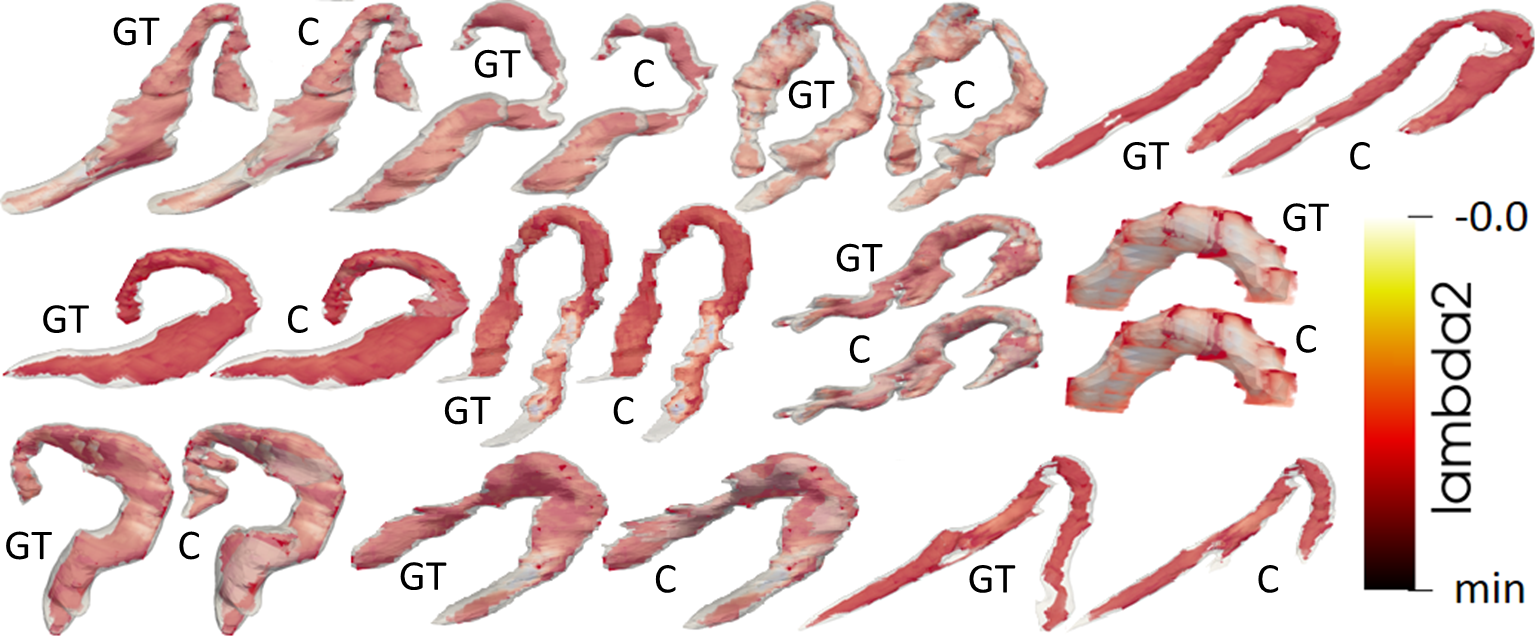}%
    \caption{\revise{This figure presents side-by-side comparisons of additional hairpin vortex cases from the channel flow dataset between the references (GT) and our results (C). The vortices are visualized with the direct volume rendering of $\lambda_2$ inside their surfaces. Previous methods were unable to extract these hairpin vortices.}}
    \label{fig:5_6A}
\end{figure}

\subsection{Results on Channel Flow}
Channel flow refers to fluid flow between two parallel flat surfaces, where both surfaces are stationary, no-slip, and impermeable. In the DNS channel flow dataset of~\cite{lee2013petascale}, both the top and bottom boundaries have the same specifications; therefore, we only present the results for the bottom boundary. Furthermore, we show results for a $256 \times 192 \times 256$ subset of the original full-domain data of $2048 \times 1536 \times 512$ grid points at frame\# 2000. The channel flow dataset has a lower grid resolution but a significantly higher Reynolds number (180 vs.\ 1000) compared to the Couette flow dataset. Figure \ref{fig:5_2A} shows zoomed-in views of several hairpin vortices extracted using our method (C) and previous methods (A, B), compared against the reference hairpin vortices. It is evident that our method extracts significantly more intact structures, whereas previous methods suffer from both under- and over-segmentation. Our method also outperforms prior approaches in terms of $F1$-score by a large margin (\cref{tab:Table2}). These results demonstrate that our approach is generalizable for identifying hairpin vortices across fluid flows with varying Reynolds numbers and grid resolutions. Additional results are provided in \cref{fig:5_6A}.

\subsection{Results on Transitional Boundary Layer Flow}
\begin{figure}[!t]
    \centering
    \includegraphics[width=0.99\linewidth]{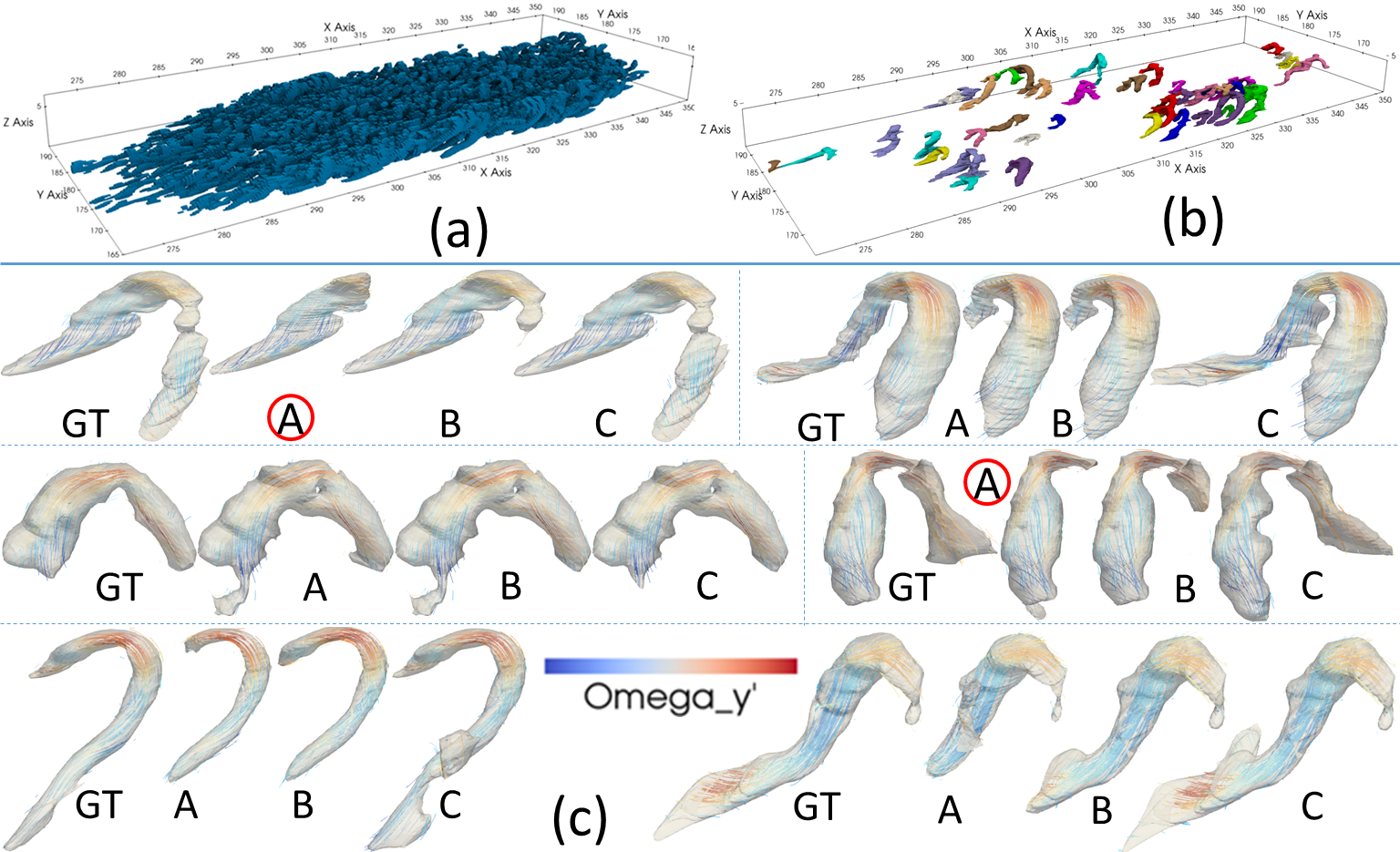}%
    \caption{\revise{This figure shows the hairpin vortices extracted from the TBL dataset~\cite{Li2008TurbulenceDB}. In (b), individual vortices are represented by distinct colors, and only the true positives are shown. 
    (c) presents a side-by-side comparison of several hairpin vortices between the references (GT), Zafar et al.~\cite{adeelhairpin2023} (A), Zafar et al.~\cite{zafar2024topological} (B), and ours (C).}}
    \label{fig:5_3A}
\end{figure}

\begin{figure}[!t]
    \centering
    \includegraphics[width=0.99\linewidth]{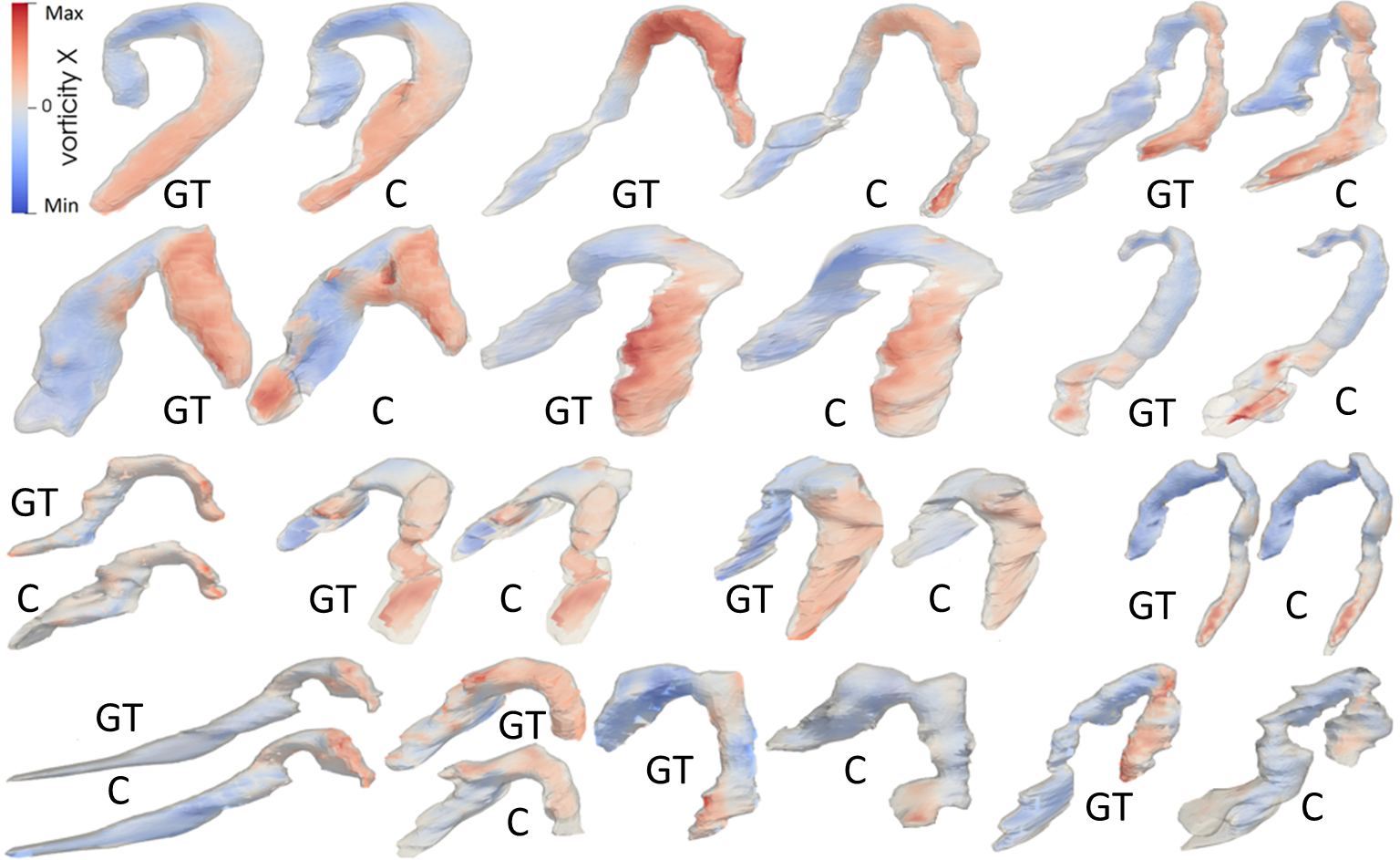}%
    \caption{\revise{This figure presents side-by-side comparisons of additional hairpin vortex cases from the TBL dataset between the references (GT) and our results (C). Color depicts the distribution of the x-component of vorticity ($\omega_x$) within the structures, highlighting the left–right counter-rotation characteristic of hairpin vortices. Previous methods were unable to extract these hairpin vortices.}}
    \label{fig:5_5A}
\end{figure}

Next, we apply our hairpin vortex identification approach to examine the transitional boundary layer (TBL) dataset from~\cite{Li2008TurbulenceDB}. In this TBL configuration, the flow originates in a laminar state at the leading edge of a stationary, no-slip, impermeable flat plate and gradually transitions to turbulence under increasing Reynolds number (i.e., Re=1502) and flow instabilities. Furthermore, we show results for a $276 \times 251 \times 224$ subset of the original full-domain data of $3320 \times 2048 \times 224$ grid points at frame\# 2350.
The statistics and qualitative comparisons for this dataset are provided in \cref{tab:Table2} and \cref{fig:5_3A}, respectively. Additional results are shown in \cref{fig:5_5A}. It is evident that as turbulence increases, our method maintains a significantly higher number of positive detections compared to the previous methods.

%% file: content/discussion.tex
\subsection{Limitations}
\label{sec:limitations}
In this section, we discuss the limitations of our method. 

\begin{figure}[!t]
    \centering
    \includegraphics[width=0.99\linewidth]{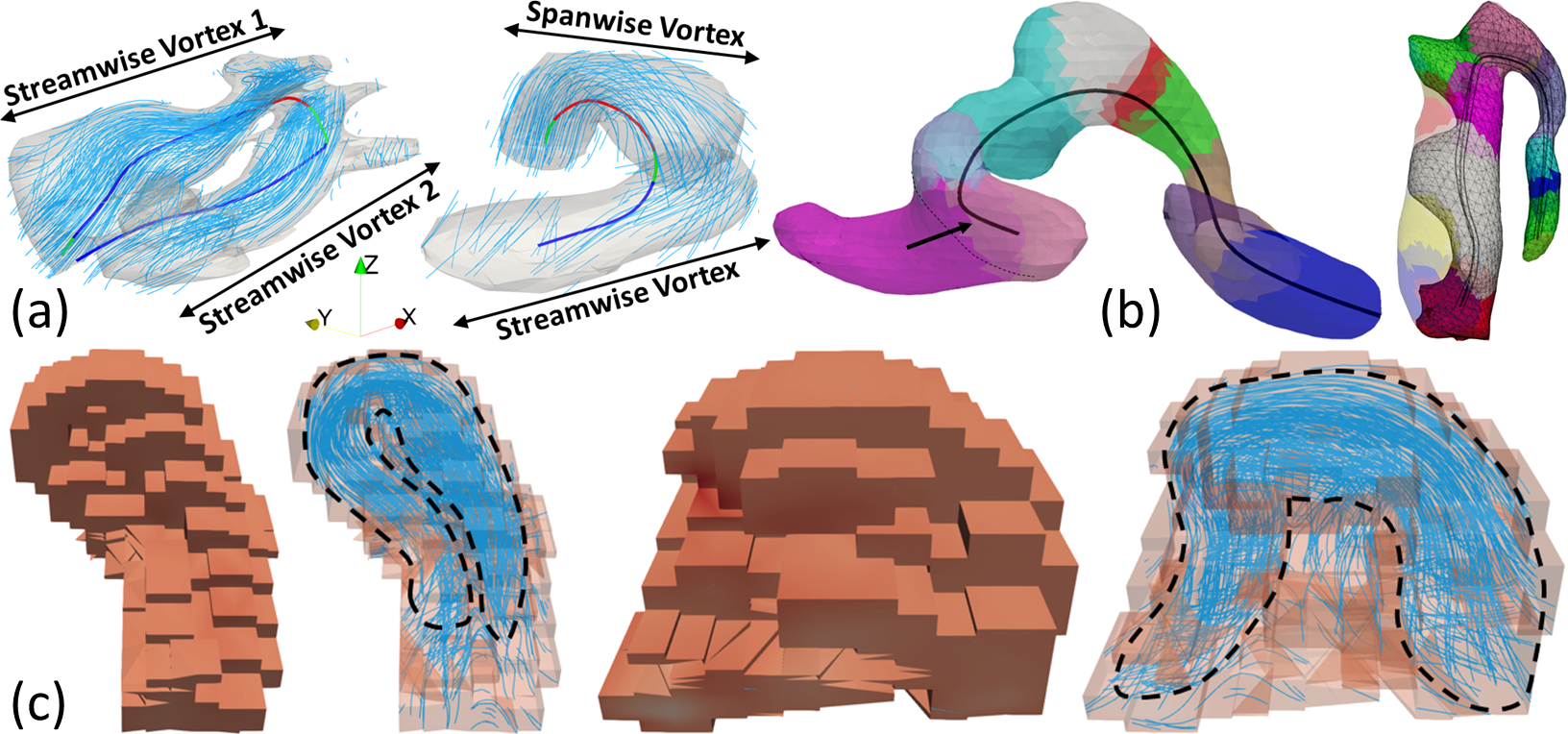}%
    \caption{(a) Examples of false positives produced by our method. Discontinuities in the vortex lines indicate that two separate vortices were incorrectly joined to form a false hairpin vortex. (b) Left: an example of under-segmentation, where an extraneous segment (pink) is incorrectly attached to a hairpin vortex. Right: an example of over-segmentation, where missing segments (shown transparent) result from vortex lines (black) failing to pass through all required segments during the rejoining process. (c) Examples of false negatives, where a hairpin vortex is embedded within a $\lambda_2$ blob that does not resemble a hairpin shape, leading to extraction of a skeleton that fails the hairpin criteria. Inspection of the internal vortex lines (blue) reveals the presence of a hairpin vortex (roughly indicated by dotted lines).}
    \label{fig:4_6}
\end{figure}

\begin{itemize}[leftmargin=*]
  \setlength{\itemsep}{1pt}
  \setlength{\parskip}{1pt}
    \item[1] \textbf{False Positives}: Segments may be incorrectly rejoined, forming a skeleton that appears to satisfy the hairpin shape criteria (\cref{fig:4_6}(a)). This typically occurs when two streamwise vortices \change{(oriented in X direction)} begin to entangle at their downstream ends (\cref{fig:4_6}(a)-left) or a spanwise vortex \change{(oriented in Y direction)} gets attached with a streamwise vortex ((\cref{fig:4_6}(a)-right)).
    \item[2] \textbf{False Negatives}: Examples of false negatives are shown in \cref{fig:4_6}(c). In these cases, a hairpin vortex is embedded within a blob of the $\lambda_2$ region that does not exhibit a clear hairpin shape. As a result, the extracted skeleton fails to meet the hairpin criteria, leading to missed detections. However, when the vortex lines (blue) inside the blob are examined, the presence of a hairpin vortex becomes evident, as roughly indicated by the dotted outlines in the figure.
    \item[3] \textbf{Under- and Over-Segmentation}: Although our method shows a significant improvement in extraction quality, as evidenced by the higher $IoU$ reported in \cref{tab:Table2}, some extraneous segments may still remain attached or points may be missed during the rejoining process, resulting in residual under- or over-segmentation as depicted in \cref{fig:4_6}(b)) and quantified in \cref{tab:Table2}. 
\end{itemize}

\section{Conclusion and Future Work}
In this work, we presented a novel method for identifying hairpin vortices. Our approach segments vortical regions in a single pass using merge tree-based segmentation followed by layering, resulting in significantly lower computational overhead than prior methods that rely on recursive segmentation and user feedback loops. By leveraging vortex lines, the method effectively rejoins relevant segments, enabling accurate hairpin vortex identification with minimal manual intervention. Comparative results show that our approach reduces both false positives and false negatives while mitigating under- and over-segmentation. Unlike clustering-based and hierarchical exploration methods, our technique streamlines detection and provides more consistent identification of hairpin vortices across diverse flow conditions.

To overcome the limitations (\cref{sec:limitations}), future work will focus on improving the rejoining process to better recover missing hairpin segments, potentially through enhanced seeding strategies for vortex line extraction. The erroneous attachment of extraneous segments can also be reduced by incorporating additional validation criteria, such as \textit{Rortex}, $Q$, or $\lambda_{ci}$, using a majority-voting scheme. Because vortices evolve through formation, interaction, and decay, some false positives and false negatives at a given time step may be resolved at later times as structures develop or separate, we plan to address this through hairpin vortex tracking. Finally, extracted hairpin vortices can be labeled to support machine learning–based approaches that replace the skeletonization step, further reducing computational cost and false positives.

\revise{For fluid dynamics studies, this method enables layer-by-layer analysis of hairpin vortices at different wall normal distances, allowing quantification of their occurrence and geometric evolution from the near-wall region to the outer layer across varying Reynolds numbers. By extracting hairpin vortices from consecutive snapshots and consistently tracking individual vortices over time, the method also enables focused analysis around low-speed streaks, revealing how vortices are arranged, how they evolve, and how often they appear together. These observations help clarify the open question of whether hairpin vortices tend to cause low-speed streak formation or whether low-speed streak formation instead leads to the formation of hairpin vortices. Moreover, the method enables physics-relevant measurements (e.g., per-structure energy content, momentum-transport proxies, and inclination/packet-angle statistics), directly linking qualitative visualization to quantitative evaluation of existing theoretical models of near-wall and outer-layer organization.}